\documentclass[
longbibliography,
nobibnotes,
notitlepage,
amsmath,amssymb,
aps,
pra,
]{revtex4-1}
\usepackage{hyperref}
\usepackage{graphicx}
\usepackage{dcolumn}
\usepackage{bm}
\usepackage{afterpage}

\begin{document}

\title{Spontaneous thermal Hall conductance in superconductors with broken time-reversal symmetry}

\author{F. Y{\i}lmaz}
\email{fyilmaz@cts.nthu.edu.tw}
\affiliation{Institute of Physics, Academia Sinica, Taipei 115, Taiwan}
\affiliation{Physics Division, National Center for Theoretical Sciences, Hsinchu 300, Taiwan}
\author{S. K. Yip}
\affiliation{Institute of Physics, Academia Sinica, Taipei 115, Taiwan}
\affiliation{Institute of Atomic and Molecular Sciences, Academia Sinica, Taipei 115, Taiwan}
\affiliation{Physics Division, National Center for Theoretical Sciences, Hsinchu 300, Taiwan}
\date{\today}

\begin{abstract}
The off-diagonal components of thermal conductance tensor, thermal Hall conductivities (THCs), have extensively been studied in recent condensed matter experiments to investigate fractionalized quantum spin liquids, and quantum Hall systems. Under zero magnetic field, THCs spontaneously become non-zero for time-reversal symmetry (TRS) broken systems, and can have contributions from topologically protected edge states. Here we focus on an additional bulk effect, impurity mechanism in TRS broken superconductors. Inspired by $Sr_2 Ru O_4$, the low temperature THC was calculated [Sup. Sci. and Tech. 29, 085006 (2016)] for the chiral p-wave superconductors induced by point impurities. Compared to topological part of THC, this contribution can be orders of magnitude larger as it scales with the density of states at the Fermi level. Motivated by TRS broken superconductors, URu$_2$Si$_2$ and SrPtAs and Sr$_2$RuO$_4$ as recently also been suggested as d-wave possibly, we calculate the THCs to $i.$ finite temperatures $ii.$ d-wave pairing states, $iii.$ finite size impurities.

For this study, the non-equilibrium quasi-classical Keldysh Green's function formalism is utilized. The THCs are calculated by the systematic expansion of the quasiclassical transport equation in the center of mass gradients, self-consistently. $\kappa_{ij}$ are obtained analytically at low temperatures ($
T \to 0$) and numerically at finite temperatures. 

We find that the impurity mechanism is dominant in $\kappa_{yx}$ at finite temperatures when compared to the topological part except at very low temperatures.There are two experimental signatures of IM on $\kappa_{yx}$: A non-monotonic temperature dependence and a sign change as a function of temperature depending on the scattering process.
\end{abstract}
\maketitle

Broken time-reversal symmetry (TRS) in the context of superconductors (SCs) is possible with a complex valued momentum dependent order parameter. In such a situation, the generalized BCS pairs can acquire a finite angular momentum. Candidate TRS broken SCs include $Sr_2 Ru O_4$ \citep{maeno2011evaluation3}, $U Pt_3$ \citep{sauls1994order5}, $U Ru_2 Si_2 $ \citep{mackenzie2003superconductivity6} and $Sr Pt As$ \citep{biswas2013evidence7} as summarized in a recent paper \citep{wysokinski2019time9}.

In addition, a typical order parameter can admit nodal points and lines (or at least a suppressed energy gap), which allows the gapless Bogoliubov quasiparticles (BQs).

Broken TRS in SCs can be investigated by a few methods. In $\mu$-spin relaxation technique \citep{amato1997heavy10}, incident spin polarized muons are precessed by local magnetic fields created by TRS broken phase of a SC. Secondly, the polar Kerr angle method \citep{yip1992circular,xia2006high11,kapitulnik2009polar} detects changes in the polarization of a polarized light incident on the surface of a superconductor. As it is proportional to the non-zero components of the electric Hall conductivity, $\sigma_{xy}$, Kerr rotation is an indication of the broken TRS. Two recent experiments measure the asymmetry in the critical field \citep{avers2018vortex} and in $dI/dV$ curves of edge states \citep{jiao2019microscopic} to detect this nature. An additional method which also constitutes the main interest of this article, is the thermal Hall conductances (THC), $\kappa_{ij}$. 

There has been strong interest in the thermal Hall coefficients, which have been discussed in magnetic systems \citep{onose2010observation,katsura2010theory,matsumoto2014thermal,cookmeyer2018spin} as well as superconductors in the vortex phase \citep{vafek2001quasiparticle,ueki2019drastic}. Recently, $\kappa_{yx}$ has received even more attention due to the search for anyonic and fractionalized excitations in condensed matter systems \citep{kitaev,moore1991nonabelions,wen1991non}, where the quantized thermal Hall conductivity has been proposed as a method to detect the topologically protected edge states in fractional quantum Hall states \citep{moore1991nonabelions,wen1991non}, Kitaev magnets \citep{kitaev,nasu2017thermal} or topological superconducting systems \citep{moore1991nonabelions,shimizuSCMajorana,nomura2012cross,sumiyoshi2013quantum}. Half-integer quantized THC has already been reported for the $\nu = 5/2$ fractional quantum Hall state \citep{banerjee2018observation2} and the $\alpha$-RuCl$_3$ Kitaev quantum spin liquid \citep{kasahara2018majorana1}.

THCs can spontaneously become non-zero for TRS broken SCs. Though the non-zero THCs can have a contribution from topologically protected edge states \citep{sumiyoshi2013quantum,imai2017thermal,yoshioka2018spontaneous13,iimura2018thermal}, {\bf here we investigate} an additional effect, the impurity mechanism (IM) for $\kappa_{yx}$ \citep{yip2016low4,arfi1989transport14}. 

Impurities are present in almost all real materials. In the normal state, they scatter the Fermi quasiparticles resulting in the decrease of the conductivities. Conventional superconductors are insensitive to non-magnetic impurities as the impurity scattering does not change the sign of s-wave gap seen by the electrons. Consequently, there are no new BQs which alter the qualitative properties of the system. Interestingly, unconventional superconductors are sensitive to even a small number of impurities \citep{hirschfeld1988consequences,hirschfeld1989electromagnetic20}. Compared to conventional superconductors at low temperatures, the formation of impurity band changes the transport properties dramatically. On one hand, impurities scatter the existing BQs and reduce the transport; on the other, they break Cooper pairs and create new BQs which enhance the thermal transport \citep{lee1993localized, graf1996electronic,shakeripour2009heat,hassinger2017vertical12}.
In this way, impurities hold a double role in thermal transport processes. 

The simplest non-trivial superconductor leading finite THCs is a p-wave superconductor. Inspired by $Sr_2 Ru O_4$, the low temperature THCs are analytically calculated \citep{yip2016low4} in the presence of point impurities. It is shown that $\kappa_{yx}$ by IM in a p-wave SC for a quasi-2D cylindrical Fermi surface with vertical line nodes is around two orders of magnitude larger than the topological contribution \citep{yoshioka2018spontaneous13}. Motivated by other candidate TRS broken superconductors, we focus on the role of IM in $\kappa_{ij}$ in d-wave superconductors. However, THCs for d-wave pairing are zero in the presence of point impurities \citep{arfi1989transport14}. In this work, hence, $\kappa_{ij}$s are calculated for $i.$finite size impurities \citep{choi1995effect,haran1998effect,kulic1999anisotropic,pisarski2003local} and $ii.$ finite temperatures (numerically) and $T\to0$ limit (analytically). 

We consider two possible irreducible representations of d-wave pairing based on the materials named above. We ${\bf find}$ that, under a temperature gradient, $\hat x \frac{-dT}{dx}$, the spontaneous $\kappa^{imp}_{yx}$ due to IM is non-zero. At finite temperatures, $\kappa^{imp}_{yx}$ is much larger than the topological contribution, $\kappa^{topo}_{yx}$ within reasonable values of the impurity concentration ($n_i$) and phase shifts ($\delta_{s},\delta_{p} \ne \{ 0, \pi/2 \}$). However, $\kappa^{imp}_{yx}$ is much smaller than the topological contribution at very low temperatures. 

{\itshape Formalism.-} THCs are calculated by using the non-equilibrium quasiclassical (QC) theory of a Fermi liquid \citep{eilenberger1968transformation,eliashberg1972inelastic,serene1983quasiclassical15}. The QC Keldysh Gfncs \citep{keldysh1965diagram16} obey the quantum transport like equation (QTE),
\begin{equation}\label{transportEq}
\big[ \varepsilon \hat{\tau}_3 - \hat{\Delta} - \check{\sigma}, \check{g} \big] + i \vec{v}_f \cdot \vec{\nabla}_{\vec{R}} \check{g} = 0, \quad \check{x} = \begin{bmatrix}
\hat{x}^R & \hat{x}^K\\
0 & \hat{x}^A \end{bmatrix}
\end{equation}
with $\check{x} \in \{ \check{g}, \check{\sigma}, \check{t}\}$ and the normalization condition $\check{g}^2 = - \pi^2$. $\varepsilon$ is the energy, $\vec{v}_f$ is the Fermi velocity, $\hat{\Delta}$ is the pairing gap function, $\check{\sigma}$ is the impurity self-energy, $\check{t}$ is the t-matrix and $\hat \tau_3 = \tau_3 \sigma_0$. Note that $R,A$ and $K$ correspond to retarded, advanced and Keldysh components. One can obtain Gfncs $\hat{g}_i^X(\hat{k},\varepsilon)$, $i \in \{0,1\}$ up to the $i$th order in the center of mass gradients. $\hat{k} =\hat k_x \hat x + \hat k_y \hat y +\hat k_z \hat z$ is the Fermi unit vector, where $\hat k_i$ is the $i^{th}$ component . In addition, $(\hat{g}^X_i)_{nm}$ denotes the $\tau_n \sigma_m$ component in the extended Nambu space. The equilibrium zeroth order retarded Gfnc is,
\begin{equation}
\label{g0R}
\hat{g}^{R}_0 (\hat k)= -\pi \frac{\varepsilon^{R} \hat{\tau}_3 - \hat{\Delta}^{R}(\hat{k})}{D^{R}(\hat{k})},
\end{equation} 
and $D^{R} = (\lvert \Delta^{R}(\hat{k})\lvert^2-(\varepsilon^{R})^2)^{1/2}$ and $\hat{\Delta}^{R}(\hat{k}) = \Re{\Delta^{R}(\hat k)}\tau_1 i \sigma_2 + \Im{\Delta^{R}(\hat k)}\tau_2 i \sigma_2$. The two irreducible representations for the d-wave pairing gap function are $\Delta^{R}_{E_{1g}}(\hat k) = \Delta_0^{R} \hat k_z (\hat k_x + i \hat k_y )$ and $\Delta^{R}_{E_{2g}}(\hat k) = \Delta_0^{R} (\hat k_x + i \hat k_y )^2$, where $\Delta_0^{R}$ is the renormalized gap size. All relations are also valid for the advanced Gfnc.

The Keldysh Gfnc, $\hat{g}^K$ can be written as a combination of the equilibrium and the anomalous parts as follows,
\begin{equation}\hat{g}^K = (\hat{g}^R -\hat{g}^A) \tanh{\frac{\varepsilon}{2 T}} + \hat{g}^K_{1a}.\end{equation}
In this work, $\tau_0 \sigma_0$ part of the anomalous (Eliashberg) Gfnc \citep{eliashberg1972inelastic}, $(\hat{g}^{K}_{1a})_{00}$ is the key function to capture the non-equilibrium effects in the current densities, $J_i(\hat{k})$. It consists of two parts, $(\hat{g}_{1a}^{K})_{00} = (\hat{g}_{1a}^{K,ns})_{00} + (\hat{g}_{1a}^{K,V})_{00}$. The first term, the non-self consistent anomalous Gfnc, $(\hat{g}_{1a}^{K,ns})_{00}$ captures the response by neglecting the non-equilibrium changes in self-energies. It has only $k_x$ component from the temperature gradient $\vec{v}_f \cdot \vec{\nabla}T$, and leads to $\kappa_{yx}=0$. The second term, the vertex correction anomalous Gfnc, $(\hat{g}_{1a}^{K,V})_{00}$ is proportional to the anomalous self-energy $\hat{\sigma}_{1a}^K(\hat k)$ (See Eq.\ref{t1aK}) with a finite $\kappa_{yx}$.

The self-energies ($\hat{\sigma}^{RAK}$), the RA Gfncs ($\hat{g}^{RA}$) are inputs to $\hat{g}^{K}$. All the equations must be solved self-consistently at each order. The details of the transport equation and the self-energy calculations are presented in the Appx.\ref{Supp1}-\ref{Supp2}, respectively. We use only the final expressions unless the underlying calculations pose a physical significance. Once the self-energies are calculated, the energy current density can be determined with the phase space sum of $(\hat{g}^{K})_{00}$.
\begin{equation}\label{currentDensity}
J_i =2 N_f \int \frac{d\hat{k}}{4 \pi} v_{f,i} \int_{-\infty}^{\infty} \frac{d \varepsilon}{4 \pi i} \varepsilon \left( \hat{g}^K \right)_{00}.
\end{equation}
In this expression, the spin degeneracy imposes the factor of two. $N_f$ is the density of states at the Fermi level and $v_{f,i}$s are the components of the Fermi velocity. It turns out, $J_i$ is non-zero only for the anomalous Gfnc, $\hat{g}^{K}_{1a}$. Moreover, $J_i$ and $\kappa_{ij}$ are related by the phenomenological relation, $J_i = \kappa_{ij} \frac{-dT}{dx_j}$.

\begin{figure}
\includegraphics[width=0.48\textwidth]{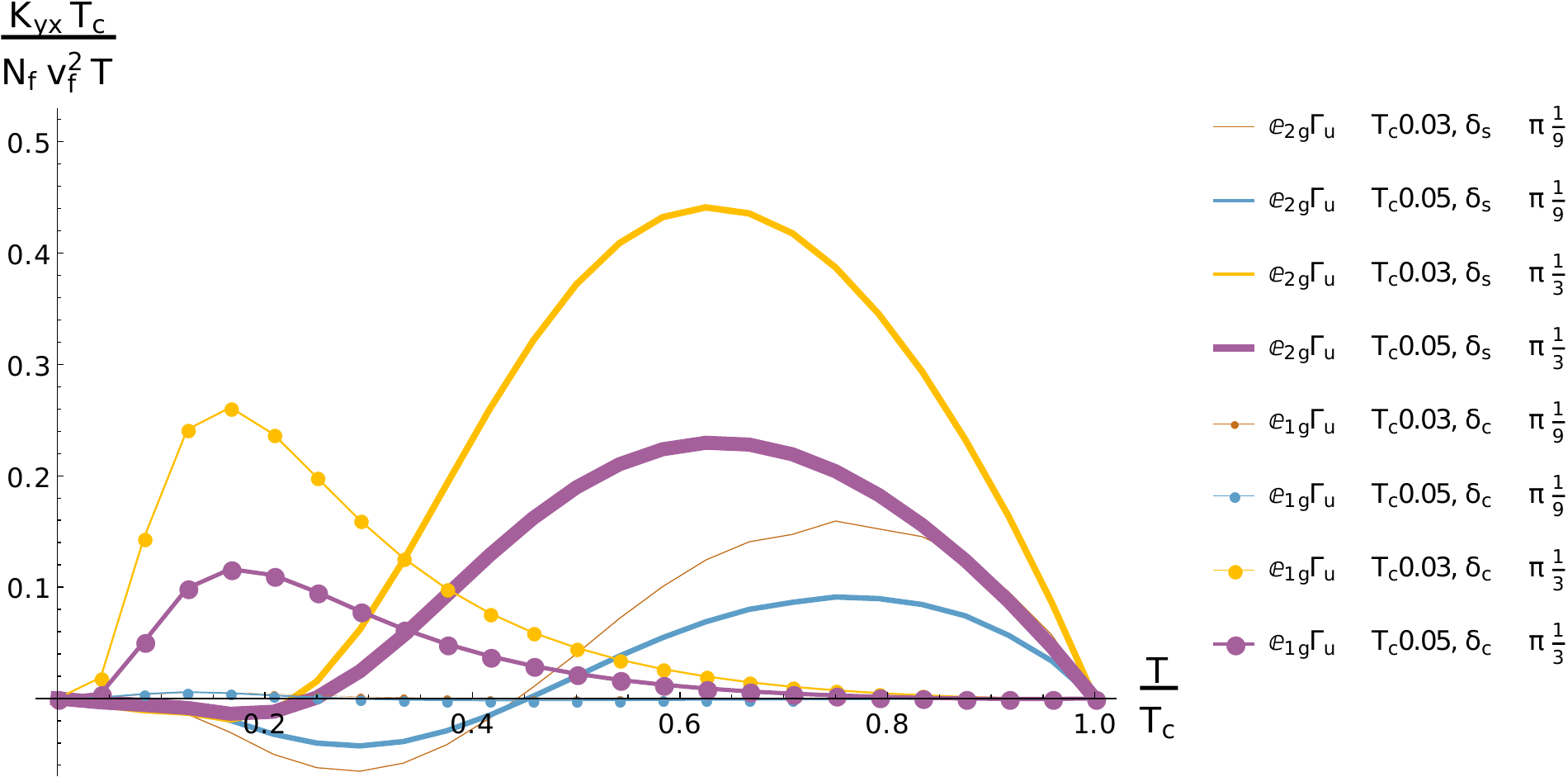}
\caption{The dimensionless $\kappa_{yx}$ as a function of temperature (in units of $T_c$). It is calculated both for $E_{1g}$(dotted lines) and for $E_{2g}$(solid lines) for two different $\delta_c$ (or $\delta_s$) and $\Gamma_u$. The s-wave scattering phase shift for $E_{1g}$ is $\delta_0 = -\pi/11$ and for $E_{2g}$ is $\delta_0 = -\pi/9$.}
\label{KyxVsTE1g}
\end{figure}

{\itshape T-matrix.-} In literature, impurities are generally modelled as points with a momentum independent scattering potential $v(\hat{k},\hat{k}') = v_0$ (in units of $T_c \times$ Volume), which greatly reduces the algebraic load. This approach is generally adequate to describe the scattering lifetime and other single particle modifications. The finite size impurities allow scattering events which depend on the momentum direction and are key to a non-zero $\kappa_{yx}$ in d-wave SCs. The first non-trivial contribution to a finite size spherical impurity is a p-wave scattering term in the scattering potential, $v(\hat{k},\hat{k}') = v_0 + 3 v_p \hat{k} \cdot \hat{k}'$. For reasons that we shall explain below, we consider a more general form,
\begin{equation}\label{scatPot}
v(\hat{k},\hat{k}')= v_0 + 3 v_s \left(\hat{k}_x \hat{k}_x' +\hat{k}_y \hat{k}_y'\right)+ 3 v_c \hat{k}_z \hat{k}_z'.\end{equation}
The scattering potential becomes spherically symmetric when $v_s = v_c$. We shall see that $v_c$ and $v_s$ play different physical roles for $\Delta_{E_{1g}}$ and $\Delta_{E_{2g}}$. The coefficients $v_0$, $v_s$ and $v_c$ are inputs to the transport equation and related to the scattering phase shifts by $\cot{\delta_{0}} = \frac{-1}{ \pi N_f v_{0}}, \cot{\delta_{s}} = \frac{-1}{ \pi N_f v_{s}}, \cot{\delta_{c}} = \frac{-1}{ \pi N_f v_{c}}$(see Supp.\ref{Supp31})

The effect of impurities are included by the t-matrix approximation,
\begin{equation}
\label{tMatrixRec}
\check{t} = v + N_f v \check{g} \check{t}.\end{equation}
The diagonal momentum components of t-matrix are proportional to the self energy as $\check{\sigma}(\hat{k}) =n_i \check{t}(\hat{k},\hat{k})$. Expanding Eq.\ref{tMatrixRec}, the Keldysh t-matrix is obtained as,
\begin{equation}
\label{tK}
\hat{t}^{K} = N_f \langle \hat{t}^{R} \hat{g}^{K} \hat{t}^{A} \rangle_{\hat{k}''} = (\hat{t}^{R} - \hat{t}^{A}) \tanh{\frac{\varepsilon}{2 T}} +\hat{t}^{K}_{1a}. \end{equation}
The recurrence relation of the anomalous t-matrix is,
\begin{equation}
\label{t1aK}
\hat{t}_{1a}^K(\hat k, \hat k') = \pi N_f \langle \hat{t}_0^R(\hat k, \hat k_1) \frac{\hat{g}_{1a}^K(\hat k_1)}{\pi} \hat{t}_0^A(\hat k_1, \hat k^{'})\rangle_{\hat k_1}.
\end{equation}
For a d-wave SC, the zeroth order Gfncs are even in $\hat k$ while the first order Gfncs are odd, e.g. $\hat{g}_{1a}^K$. Then, a non-zero average in Eq.\ref{t1aK} is possible when even-odd combinations of $\hat t_0^R,\hat t_0^A$ are matched. However, a finite anomalous self energy, $\hat{\sigma}_{1a}^K (\hat k)= n_i \hat{t}_{1a}^K (\hat k, \hat k)$ does not automatically imply a finite $\kappa_{yx}$, as it must also have an odd momentum component in $\hat y$ direction.

\subsection{$E_{1g}$ Order Parameter, $\Delta_k = \Delta_0 \cos{\theta} \sin{\theta} e^{i \phi}$}
The energy gap function $\Delta_{E_{1g}}(\hat k)$ admits a nodal line around the equator and two point nodes at both poles of the Fermi level. For the impurity scattering potential in Eq.\ref{scatPot}, one can argue that the BQs around the equator not only are much larger in numbers but also have much larger transverse momentum components, which would lead to a greater $\kappa_{yx}$ compared to the BQs around the poles. Naturally, the main contribution to new BQs around the equator should be dominated by an effective potential, $v(\hat k, \hat k') \to v_0 + 3 v_s (\hat k_x \hat k_x' +\hat k_y \hat k_y')$. However, in this limit, $\hat{g}^{K,V}_{1a}$ becomes an odd function of $k_x$ only. Without an odd $k_y$ part in $\left(\hat{g}^{K,V}\right)_{00}$, the angular part of the integration in Eq.\ref{currentDensity} vanishes, $J_y \sim \int \frac{d\Omega}{4\pi} v_{f,y} k_x (...) \sim 0$, where $(...)$ includes the rest of the irrelevant terms.  Therefore, a finite $v_c$ is necessary for a non-zero $\kappa_{yx}$.

For the sake of clarity, we focus on non-zero $v_c$ case, $v(\hat k, \hat k') = v_0 + 3 v_c \hat k_z \hat k_z'$ (see Supp. \ref{Supp321} for discussion where both $v_s$ and $v_c$ are finite). Using Eq.\ref{tMatrixRec}, the corresponding t-matrix is calculated as
\begin{equation}\label{t0R}
\hat t^R_0(\hat k, \hat k')= \hat t_{00}^R + \hat t_{cc}^R \frac{\hat k_z \hat k_z'}{k_f^2}=\frac{-1}{\pi N_f} \left( \frac{\cot \delta_0 +i \tilde \gamma_{0} \hat \tau_3}{C_{0}} + 3 \frac{\cot \delta_c + i \tilde \gamma_{c} \hat \tau_3}{C_{c}} \hat k_z \hat k_z' \right) \nonumber
\end{equation}
where $C_{\nu} = \cot^2 \delta_{\nu} + (\tilde{\gamma}_{\nu})^2$ with $\nu \in \{0, c,s \}$. $\tilde{\gamma}_0 = \langle \frac{-i \varepsilon^R}{D^R} \rangle$, $\tilde{\gamma}_c = 3 \langle \frac{-i \varepsilon^R }{D^R} \hat k_z \hat k_z' \rangle$ are the momentum averages of $(\hat{g}_0^R)_{30}$. $\hat t_0^R$ is a diagonal matrix in the particle-hole space and does not contain particle$\to$hole type scattering events. The s-wave scattering part of the t-matrix, $\hat t_{00}^R$ connects all incoming ($\hat k'$) and the outgoing ($\hat k$) particles (holes) with equal probabilities, whereas $\hat t_{cc}^R$ favours the scattering of BQs between the poles, $N \to N, N \to S, S \to N, S \to S$.

The zeroth order self-energy, $\hat{\sigma}_{0}^R (\hat k)= n_i \hat{t}_{0}^R (\hat k, \hat k)$ is plugged back into Eq.\ref{transportEq} and $\hat g_0^R$ is obtained self-consistently (for the self-consistency relations and the modified density of states see see Appx. Eq.\ref{selfConsis} and Fig. \ref{ModifiedDOS}). Once $\hat g_0^R$ and $\hat \sigma_0^R$ are determined, the advanced components are obtained by the general symmetry relations, $\hat x^A = \hat \tau_3 (\hat x^R)^\dagger \hat \tau_3$\citep{serene1983quasiclassical15}, where $\hat x \in \{ \hat \sigma, \hat g \}$ and $\hat{t}^A(\hat k, \hat k') = \hat \tau_3 \left(\hat{t}^R(\hat k', \hat k) \right)^\dagger \hat \tau_3$. 

The non-self consistent part of $(\hat{g}_{1a}^{K})_{00}$ is explicitly found as, 
\begin{equation}\label{g1akns00}
\frac{(\hat{g}_{1a}^{K,ns})_{00}}{ \pi b(\varepsilon,T)}= \frac{2 \Re{D^R}}{D_1} \left(1+ \frac{ \lvert \varepsilon^R \rvert^2 - \lvert \Delta(\hat{k}) \rvert^2}{\lvert D^R \rvert^2} \right) \hat k_x,
\end{equation} 
where $D_1= 4 [(\Re{D^R})^2 - (\Im{(\sigma^R_0)_{00}})^2]$ and $b(\varepsilon,T) = i v_f \frac{ -d T}{dx} \frac{ \varepsilon}{2 T^2} sech^2{\frac{\varepsilon}{2 T}}$.

Plugging $\hat{g}_{1a}^{K,ns}$ (see Appx. Eq.\ref{g1aKnsSupp} for the full form) and $\hat{t}^{R,A}$ into Eq.\ref{t1aK}, the anomalous self energy is determined as $\hat{\sigma}_{1a}^{K,ns} = \Gamma_u \hat{G} \tau_1 i \sigma_2 b(\varepsilon,T) \hat{k}_z$ where $\hat G=G_0 + i G_3 \tau_3$ with real $G_0$, $G_3$ and $\Gamma_u=n_i/\pi N_f$. It is the non-self-consistent solution when only $\hat g_{1a}^{K,ns}$ is used. Using $\hat g_{1a}^{K,V} = \frac{2 \Re{D^R} \hat{g}_0^R/(- \pi) +2 i \Im{(\sigma^R_0)_{00}}}{D_1} (\hat{g}^{R}_0 \hat{\sigma}^{K}_{1a}- \hat{\sigma}^{K}_{1a} \hat{g}^{A}_0)$, we can calculate $\hat g_{1a}^{K,V}(\hat k,\varepsilon)$ and when plugged back into Eq.\ref{t1aK}, it reveals the self-consistency condition that renormalizes $\hat G \to \hat{\tilde{ G}}$. We find $\hat g_{1a}^{K,V}(\hat k,\varepsilon)$ as 
\begin{equation}\label{g1aKVexp}
\frac{\hat g_{1a}^{K,V}(\hat k,\varepsilon)}{\pi \Gamma_u b(\varepsilon,T)} = \hat{\tilde{X}}_c\hat{k_x} + \hat{\tilde{X}}_s \hat{k_y} + \hat{\tilde{Y}}(\hat{k}_x + i \hat k_y \tau_3) \tau_1 i \hat \sigma_2.
\end{equation}
where $ \hat{\tilde{X}}_i = \tilde{X}_{i0}+\tilde{X}_{i3}\tau_3$, with $i\in\{c,s\}$ and $\hat{\tilde{Y}} = \tilde{Y}_{0}+ i \tilde{Y}_{3}\tau_3$ are lengthy expressions given in see Appx. Eqs.\ref{g1aKVSuppE1g}-\ref{YcoeffE1g}. The components of $(\hat{g}_{1a}^{K,V})_{00}$ along $k_x$ and $k_y$, $\tilde{X}_{c,0}(\varepsilon,\theta)$ and $\tilde{X}_{s,0}(\varepsilon,\theta)$ are real, even functions of $\varepsilon-\cos \theta$ and finite only when $\Delta_0 \ne 0$. Therefore, the vertex correction vanishes in the absence of superconductivity. {\bf More generally}, all the contributions to the thermal Hall current comes from the average, $\langle (...)_{\hat k} \hat{\Delta}(\hat k) \vec{v}_f \cdot \vec{\nabla} T \rangle$ appearing in $\hat \sigma_{1a}^K$ of Eq.\ref{t1aK}, where $(...)_{\hat k}$ is odd in $\hat k$ and represents the rest of the terms. The average is non-zero when $ (...)_{\hat k}$ have components along both $\vec{\nabla} T$ and $\hat{\Delta}(\hat k)$. 

The non-self consistent Gfnc, $\hat g_{1a}^{K,ns}$ creates only a finite longitudinal current, $J_x^{ns} = N_f v_f^2 \frac{-dT}{dx} \int_{-\infty}^{\infty} d \varepsilon \frac{ \varepsilon^2}{2 T^2} sech^2{\frac{\varepsilon}{2T}} \Omega_0(\varepsilon)$, and $\kappa_{xx}^{ns}$ is found as
\begin{eqnarray}\label{JxnsE1g}
\frac{\kappa^{ns}_{xx} T_c}{\frac{\pi^2}{3} N_f v_f^2 T} &=& \frac{6 T_c }{\pi^2} \int_{-\infty}^{\infty} \frac{d \tau \tau^2}{ \cosh^2(\tau)} \Omega_0(2 \tau T ),
\end{eqnarray}
$\tau = \varepsilon/2T$ is the dimensionless energy, $T_c$ is the critical temperature and $\Omega_0(\varepsilon) = \langle \frac{4 \Re{D^R}}{D_1} \left(1+ \frac{ \lvert \varepsilon^R \rvert^2 - \lvert \Delta \rvert^2}{\lvert D^R \rvert^2} \right) \hat k_x^2$. The average $\Omega_0$ is in units of $\sim \frac{1}{E}$.

The vertex correction Keldysh Gfnc, $\hat{g}_{1a}^{K,V}$ leads the following THCs
\begin{eqnarray}\label{KyxKxx}
\frac{\begin{bmatrix}\kappa^V_{xx}\\\kappa_{yx}\end{bmatrix} T_c}{\frac{\pi^2}{3} N_f v_f^2 T} &=& \frac{6 \Gamma_u T_c}{\pi^2} \int_{-\infty}^{\infty} \frac{d \tau \tau^2}{ \cosh^2(\tau)} \langle \hat k_x^2\begin{bmatrix} \tilde{X}_{c0}(2 \tau T,\theta)\\ \tilde{X}_{s0}(2 \tau T,\theta) \end{bmatrix} \rangle_{\theta}\nonumber \\ .
\end{eqnarray}
where the both sides of Eq.\ref{JxnsE1g} and Eq.\ref{KyxKxx} are dimensionless. 

The expressions obtained in Eq.\ref{JxnsE1g} and Eq.\ref{KyxKxx} are general where the explicit forms are investigated in the low temperature limit (for the finite temperature analysis see Appx. \ref{Supp322} and \ref{Supp332}). For $E_{1g}$ case in the low temperature limit, the BQs populates only the vicinity of Fermi level as $\varepsilon \to 0$ and $(\hat{g}_0^R)_{30}$ remains only the impurity bandwidth, $\varepsilon^{R,A} \to \pm i \gamma(\theta) = i (\gamma_{0} + \gamma_{c} \hat k_z^2)$ with $\gamma_0, \gamma_c$ being positive definite. With the help of explicit $\tau$ integration in Eq.\ref{JxnsE1g}, $ \frac{6}{\pi^2}\int d\tau \tau^2 sech^2 \tau = 1$, the energy integral leads to $\frac{6}{\pi^2} \int_{-\infty}^{\infty} d \tau (...) \to \Omega_0(0) =2 \langle \frac{\gamma^2 }{D^3} \hat k_x^2 \rangle_{\theta} \sim 1/\Delta_0$, where $D = \left( \lvert \Delta (\hat k) \rvert^2 + \gamma^2 \right)^{1/2} $. Then, the diagonal current is found as $\frac{\kappa^{ns}_{xx} T_c}{\frac{\pi^2}{3} N_f v_f^2 T} \sim T_c /\Delta_0$ \citep{graf1996electronic} with a universal value. The vertex correction coefficients, $\kappa_{ji}^V$s become,
\begin{equation}
\label{kyxE1gT0approx}
\frac{\begin{bmatrix}
\kappa^V_{xx}\\ \kappa_{yx}\end{bmatrix} T_c}{\frac{ \pi^2 }{3}N_fv^2_f T} \approx \Gamma_u T_c \frac{4 \beta_3^2(0)}{C_0 C_c Det}\begin{bmatrix}\cot \delta_0 \cot\delta_c - \tilde{\gamma}_{0}\tilde{\gamma}_{c}\\ \cot \delta_0 \tilde{\gamma}_{c} +\cot\delta_c \tilde{\gamma}_{0}\end{bmatrix}.
\end{equation}
The integral average is $\beta_3=2 \langle \frac{ \gamma(\theta) \Delta_0}{D^3}  \hat k_z^2 \hat k_x^2 \rangle$ is finite for $\Delta(\hat k)^{E_{1g}}\propto \hat k_z (\hat k_x + i \hat k_y) $ due to the factor of $k_z$ coming from the potential $v_c$ and the factor of $k_x$ from the temperature gradient and evaluated as $\beta_3(0)  \sim \frac{\gamma_0}{\Delta_0^2}\ln{\frac{\Delta_0}{\gamma_0}}$ as $T\to 0$. Also, $\tilde{\gamma}_0 = \langle \frac{ \gamma(\theta) }{D}\rangle \sim \frac{\gamma_0}{\Delta_0}\ln{\frac{\Delta_0}{\gamma_0}}$, $\tilde{\gamma}_c = 3\langle \frac{ \gamma(\theta) }{D} \hat k_z^2 \rangle \sim \frac{\gamma_0}{\Delta_0}$ (see Appx. Eq.\ref{vertexCorrSelfE1g} for $Det$). Eq.\ref{kyxE1gT0approx} shows that both $\delta_0$ and $\delta_c$ must be non-zero for a finite $\kappa_{yx}$. Interestingly, even in the unitary limit of s-wave scattering phase shift $\delta_0 = \pi/2$, this is achieved for $\delta_c \in (0, \pi/2)$, and vice versa. The vertex correction  coefficients in Eq.\ref{kyxE1gT0approx} are evaluated as
\begin{equation}
\label{kyxE1gT0approxOrder}
\sim \frac{T_c}{\Delta_0} \frac{\gamma_0^2  \ln{\frac{\Delta_0}{\gamma_0}}}{\Delta^2_0}  \begin{bmatrix} \cot \delta_0 \cot \delta_c -  \frac{\gamma^2_0}{\Delta^2_0} \ln{\frac{\Delta_0}{\gamma_0}}\\ \frac{\gamma_0}{\Delta_0}\left( \cot \delta_0 +  \cot \delta_c \ln{\frac{\Delta_0}{\gamma_0}}\right) \end{bmatrix}\frac{1}{C_c}.
\end{equation}
For non-zero and finite values of  $\cot \delta_0$ and $ \cot \delta_c$, the sign of $\kappa_{yx}$ and $\cot \delta_c$ are same. In addition, the unitless form of $\kappa_{yx}$ is significantly a small number.

At finite temperatures, in Fig.\ref{KyxVsTE1g}, we numerically calculate $\kappa_{yx}$ (dotted lines and in units of $\frac{N_f v_f^2 T}{T_c}$) as a function of temperature for two sets of phase shifts $\{ (\delta_0, \delta_c)\}\in\{ (-\frac{\pi}{9},\frac{\pi}{9}),(-\frac{\pi}{9},\frac{\pi}{3})\}$ and $\Gamma_u \in \{0.03, 0.05 \} T_c$ (for $\kappa_{yx} $ vs. $\Gamma_u$ relation see Appx.\ref{Supp4}). It peaks within the range $0.1-0.8$ $T_c$. Similar to this finding, the THCs were predicted previously for p-wave pairing to have its maximum value around $T \sim \Delta_0/2$ \citep{arfi1989transport14}. In our case, there are several contributions to $\kappa_{yx} $ in competition which can lead opposite directions. Depending on the temperature and the phase shifts, $\kappa_{yx} $ can change sign (for full $\kappa_{yx} $ in the space of $\delta_0-\delta_c$ see Fig.\ref{Kyxd0dsE2gE1g}). 

\subsection{$E_{2g}$ Order Parameter, $\Delta_k = \Delta_0 \sin^2{\theta} e^{i 2\phi}$}
The energy gap function $\Delta_{E_{2g}}$ has two point nodes at the poles. Even if $v_c$ part of $v(\hat k, \hat k')$ in Eq.\ref{scatPot} is dominant for BQs around the poles, it can again be shown that it generates no thermal current in $y$-direction. It is because the integral average in Eq.\ref{t1aK} vanishes due to the orthogonal components in momentum space, $\hat{\sigma}^{K}_{1a} \sim \langle (...)_{even} k_x k_z \rangle \sim 0$, then $\kappa_{yx} \to 0$. Note that $ (...)_{even}$ corresponds to the rest of the irrelevant terms in Eq.\ref{t1aK} (see Supp.\ref{Supp332}). Therefore, we focus on the equator scattering potential $v(\hat k,\hat k') = v_0 + 3 v_s (\hat k_x \hat k_x' + \hat k_y \hat k_y'$). In this limit, the t-matrix is
\begin{eqnarray}\label{t0RE2g}
\hat{t}_0^R(\hat k, \hat k') &=& \hat{t}^{R}_{00} + \Big(\hat{t}^{R}_{ssc} [ \cos{(\phi-\phi') }- \frac{1}{2 } \alpha e^{i (\phi - \phi') \hat \tau_3}] \nonumber- \frac{3 \alpha_0}{2 \pi N_f} e^{i (\phi + \phi') \hat \tau_3}\hat \tau_1 i \hat \sigma_2\Big) \sin \theta \sin \theta',\\
\hat{t}^{R}_{ssc}&=&\frac{-3}{\pi N_f} \frac{\cot \delta_s + i \tilde \gamma_{s} \hat \tau_3}{C_{s}}.
\end{eqnarray}
Note $\hat{t}^{R}_{00}$ has the same form with Eq.\ref{t0R}$, \alpha_0 = \tilde{\gamma}^1/C_{s2}$, $\alpha = \tilde{\gamma}^1 \alpha_0$, $C_{s2} = \cot^2{\delta_s}+ (\tilde{\gamma}_{s})^2 + (\tilde{\gamma}^1)^2$ where $\tilde{\gamma}^1=\langle \frac{3}{2} \frac{\tilde{\Delta}_0 \sin^4 \theta}{D^R} \rangle$ and $\tilde{\gamma}_s=\langle \frac{3}{2} \frac{-i\varepsilon^R \sin^2 \theta}{D^R} \rangle$. $\tilde{\Delta}_0 =\frac{\Delta_0}{d}$ is the renormalized gap size with $d = \lvert 1 + \frac{3}{2} \Gamma_u \frac{\tilde{\gamma}^1}{C_{s2}} \rvert^2$. For the sake of simplicity, $\Delta_0^{R,A}$ is considered as real and equal to avoid the sufficiently small $\pm$ imaginary parts arising in the retarded and the advanced components. In Eq.\ref{t0RE2g}, the term with the coefficient $\alpha$ clearly shows the asymmetry in the particle-hole space. It is a spontaneously generated skew-scattering effect \citep{skewScatteringAHE} not present in $E_{1g}$ case (unless both $v_s$ and $v_c$ are included) and leads an additional contribution to $\kappa_{yx}$ in the superconducting state ($\alpha \propto \tilde{\gamma}^1 \ne 0$). 

As discussed in the paragraph following Eq.\ref{g1akns00}, the other sources of $\kappa_{yx}$ are the non-zero averages from $\langle (...)_{\hat k} \hat{\Delta}(\hat k) v_{f,y} \rangle$ in Eq.\ref{currentDensity}. Each term contributes to the $\kappa_{yx}$ through the anomalous self-energy $\hat \sigma_{1a}^K$. Note that $\hat g_{1a}^{K,ns}$ and $J^{ns}_x $ have the same form as in the $E_{1g}$ case (see Eq.\ref{g1akns00} and Eq.\ref{JxnsE1g}) and $\hat \sigma_{1a}^K$ can be obtained though acquires a more complicated form. Plugging $\hat\sigma_{1a}^{K,ns}(\hat k,\varepsilon)$ into $\hat g_{1a}^{K,V}(\hat k,\varepsilon)$, one can see that the anomalous vertex Gfnc has the same form with Eq. \ref{g1aKVexp}. The only difference lies in the explicit expressions of $\hat X_c, \hat X_s$ and $\hat Y$ (see Appx. Eq.\ref{g1aKVCoeffE2g}). In Fig.\ref{KyxVsTE1g}, $\kappa_{yx}$ (in units of $\frac{N_f v_f^2 T}{T_c}$) is numerically calculated for two different pairs of phase shifts for the impurity concentration $\Gamma_u = 0.04 T_c$. The center of the peak for $\kappa_{yx}$ varies as a function of phase shifts and changes sign at low temperatures. 

There are two groups of contributions to $\kappa_{yx}$ with opposite signs. At low temperatures, the first group (Eq.\ref{E2gLowTIntegrals}) is finite and in Fig.\ref{KyxVsTE1g} leads $\kappa_{yx}\le0$ for the given phase shifts within. In addition, around $T\sim \Delta_0/2$, the second group (Eq.\ref{E2gFiniteTIntegrals}) dominates the heat current as it is proportional to $(\Im{\left( \hat\sigma_0^R\right)_{00}})^2 \ge 0$ where the sign is reversed, $\kappa_{yx}\ge 0$. Beware that non-zero $\Im{\left( \hat\sigma_0^R\right)_{00}}$ creates an asymmetry in the lifetime for electrons and holes (for details see below Appx. Eq.\ref{kyxE2gT0approxSupp}).

In the low temperature limit, $\varepsilon^R \to i(\gamma_0 + \gamma_s \sin^2 \theta)$, where $\gamma_0$ and $\gamma_s$ are positive definite bandwidths. $\kappa^{ns}_{xx}/\left( \frac{\pi^2}{3}N_f v_f^2 \frac{ T}{T_c} \right) = \Omega_0 \sim \frac{T_c}{\Delta_0}\frac{\gamma_0}{\tilde{\Delta}_0}$. The explicit vertex corrections to THCs are lengthy (see Appx. Eq.\ref{kyxE2gT0Supp}), but for non-vanishing and non-divergent values of $\cot \delta_0$ and $\cot \delta_s$, we can estimate the magnitudes in Eq.\ref{kyxE1gT0approxOrder} as,
\begin{eqnarray}\label{kyxE2gT0approx}
&\sim& \frac{T_c}{\Delta_0}\frac{\gamma_0}{\tilde{\Delta}_0} \ln{\frac{\tilde{\Delta}_0}{\gamma_0}}\Re{\alpha}  \begin{bmatrix} - \frac{\cot \delta_s}{C_s} \\ 1\end{bmatrix} \cot \delta_0
\end{eqnarray}
Note that $\Re{\alpha} \sim \frac{1}{\cot^2 \delta_s + (\tilde{\gamma}^1)^2 }> 0$. The sign of $\kappa_{yx}$ and $\cot \delta_0$ are same and the unitless form of $\kappa_{yx}$ is again a very small number. 

In summary, we find that even a very small anisotropic phase shift can lead to very large THC at finite temperatures. Noting the phase shift dependence of $\kappa^{imp}_{yx}$, the unit that is used to non-dimensionalize it, $N_f v_f^2 $ can be re-expressed as $\sim \frac{E_f}{\tilde{\Delta}_0} k_f$, where typically $\frac{E_f}{\tilde{\Delta}_0} \sim 10^2-10^3$. Therefore, $\kappa^{imp}_{yx}$ can be an order of magnitude larger than the topological contribution $\kappa^{topo}_{yx}$ which is of the order $k_f$\citep{yoshioka2018spontaneous13} (Boltzmann constant, $k_B =1$), except for very low temperatures.  There is also one allocated section in the Appx.\ref{Supp4} for the discussion of the relation between $\kappa_{yx}$ and the impurity concentration.

Our results are not only specific to superconductors but should be considered as an example of a more general effect in TRS broken systems which host the emergent excitations as the heat carriers. The impurity generated contribution to THCs due to emergent excitations such as broken BCS pairs, phonons, magnons and fractional excitations could be significantly important along with the topological contribution. We believe this point of view could enrich the discussions on the recent thermal Hall conductance measurements cited above. It still requires further investigations to pose a quantitative universality of such a result.

\begin{acknowledgments}
F.Y. and S.K.Y. acknowledge the Ministry of Science and Technology of Taiwan with grant number 107-2112-M001-035-MY3.
\end{acknowledgments}
\bibliography{KyxTRSBrokenSup_PRL_v18}
\setcounter{equation}{0}
\setcounter{figure}{0}
\setcounter{table}{0}
\makeatletter
\renewcommand{\theequation}{A\arabic{equation}}
\renewcommand{\thefigure}{A\arabic{figure}}
\appendix
\section{Quasi-classical Quantum Transport Equation}\label{Supp1}

The quasi-classical approach is an effective description of the dynamics of fermions. It is a renormalized and linearized theory \citep{serene1983quasiclassical15}, which is is capable of describing both the static and dynamical properties of quasiparticles. After $\xi_k$ integration, the quasi-classical Green's functions (Gfncs) are obtained with the following $\tau_3$ convention, $\check{g} \sim \int_{\xi_k} \check{\tau_3} \check{G}$, where $\check{G}$ is the full Gfnc in the Keldysh space. $\check{g}$ is guided by a quantum kinetic equations similar to classical systems and also it inherits microscopic properties such as spin, electron and holes within a classical transport formalism. In this respect, it is much easier to obtain system observables including density, currents, magnetization etc. In addition, the Keldsyh QC Gfncs expand the capability of this theory to non-equilibrium behaviour. In our case, a temperature gradient generates a non-equilibrium bulk thermal Hall current. The $\kappa_{ij}$s are obtained by a systematic expansion of the QTE in the gradients of the slowly varying bulk center of mass coordinate (CoM), $\vec{R}$. The QC method can correctly address the new energy scale due to the formation of impurity bands even at low temperatures and allows for non-zero THCs meanwhile the Boltzmann kinetic equation approach fails to describe \citep{arfi1989transport14}.

The quasi-classical Gfncs obey the equation transport like equation as follows,
\begin{equation}\label{transportEqu}
\big[ \varepsilon \hat{\tau}_3 - \hat{\Delta} - \check{\sigma}, \check{g} \big] + i \vec{v}_f \cdot \vec{\nabla}_{\vec{R}} \check{g} = 0 \end{equation} with the normalization condition $\check{g}^2 = - \pi^2$. There are two d-wave pairing gap functions considered in this work, $\Delta^{E_g}_k = \Delta_0(T) \frac{k_z (k_x + i k_y)}{k^2_f}$ and $\Delta^{E_2g}_k = \Delta_0(T)\frac{ (k_x + i k_y)^2}{k^2_f}$. 

In the presence of the impurities, the translation invariance can be re-established by the impurity averaging. The effects of impurities are introduced through the impurity self-energy term by a t-matrix approach. Starting from the clean equilibrium limit, one can at each order obtain the RAK components of Gfncs, $\hat{g}^K(\hat{k},\vec{R},\varepsilon,T)$. Let us expand all terms up to the first order in the CoM gradients, $\vec{\nabla}_{\vec{R}}$. 
\begin{eqnarray}
\check{g} &=& \check{g}_0 + \check{g}_1,\nonumber \\
\hat{\Delta} &=& \hat{\Delta}_0 + \hat{\Delta}_1,\\
\check{\sigma} &=& \check{\sigma}_0 + \check{\sigma}_1.\nonumber
\end{eqnarray}
$\hat \Delta (\hat k)$ is self-consistently determined by with the following sum. 
\begin{equation}\label{pairingSelfCon}
\Delta (\hat k) = \int \frac{d \hat k'}{4 \pi} \mathcal{V}(\hat k, \hat k') f^K(\hat k').
\end{equation}
Note that the first order off-diagonal component of Keldysh propagator, $f^K(\hat k')$ is odd in $\hat k'$ and $\mathcal{V}(\hat k, \hat k')$ is even, therefore $\hat{\Delta}_1 = 0$. In addition, for the practical reasons, we consider a model for the maximum gap size $\Delta_0(T)$ as in [2]

Plugging the expanded terms into the Eq.\ref{transportEqu}, the equations for $RAK$ components at each order can be obtained. The zeroth order transport equations for $R,A$ Gfncs,
\begin{equation}\label{0thOrderTransportEq}
\big[ \varepsilon \hat{\tau}_3 - \hat{\Delta}_0 - \hat{\sigma}_0, \hat{g}^{R,A}_0 \big] = 0, (\hat{g}^{R,A})^2 = - \pi^2.
\end{equation}
$\hat\tau_0$ component of $\hat{\sigma}_0$ commutes with $\hat{g}^{R,A}$ and it should be dropped in the zeroth order. The zeroth order RA Gfncs are found as
$$\hat{g}^{R,A}_0 = -\pi \frac{\varepsilon^{R,A} \hat{\tau}_3 - \hat{\Delta}^{R,A}_0(\hat{k})}{D^{R,A}(\hat{k})},$$ 
with a normalization condition $(\hat{g}^{R,A}_0)^2 = - \pi^2$ and $D^{R,A} = \sqrt{\lvert \Delta^{R,A}(\hat{k})\rvert^2-(\varepsilon^{R,A})^2}$. The effect of impurity scattering process is introduced through the self-energy term, $\hat{\sigma}^R(\hat{k}$. It modifies the elements of the equilibrium Gfncs as $\varepsilon^{R} = \varepsilon - (\sigma^{R}_0)_{30}$ 
and $\hat \Delta^{R} = \Delta (\hat{k}) + (\sigma^{R}_0)_{off-diag}$. 

Expanding the Keldysh component of Eq.\ref{transportEqu}, the zeroth order Keldysh Gfnc can be expressed as $\hat{g}^K_0 = (\hat{g}^{R}_0-\hat{g}^{A}_0) \tanh{\frac{\varepsilon}{2 T}}$.
The retarded and the advanced components have the information on the spectral density of the system while the Keldysh component reveals how these states are occupied.

Eq.\ref{transportEqu} in the first order for $R,A$ are
\begin{equation}
\hat{g}^{R,A}_1 (\hat{k},\varepsilon) = -\frac{\hat{M}^{R,A}}{2 (D^{R,A})^2} \Big( i \vec{v}_f \cdot \vec{\nabla}_{\vec{R}} \hat{g}_0^{R,A} - \left[ \hat{\sigma}_1^{R,A} , \hat{g}_0^{R,A} \right] \Big)
\end{equation}
We do not have to calculate $\hat{g}^{R,A}_1$ explicitly as the main concern of this article is to obtain the first order Keldysh Gfnc. $\hat{g}_1^K$ has the equilibrium and the anomalous parts as follow,
$$\hat{g}_1^K = (\hat{g}^R_1 -\hat{g}^A_1) \tanh{\frac{\varepsilon}{2 T}} + \hat{g}^K_{1a}.$$
The anomalous Gfnc (Eliashberg propagator), $\hat{g}^{K}_{1a}$ is the key function for the current densities $J_i(\hat{k})$ and consists of two parts, $\hat{g}^K_{1a} = \hat{g}^{K,ns}_{1a}+\hat{g}^{K,V}_{1a}$. The first term is
\begin{equation}\label{g1aKnsSupp}
\hat{g}^{K,ns}_1 (\hat{k},\varepsilon) = \hat{N}^R (\hat{g}^{R}_0-\hat{g}^{A}_0) \frac{k_x}{k_f} b(\varepsilon,T),
\end{equation}
where, 
\begin{equation}\label{NR}
\hat{N}^R = \frac{(D^R + D^A) \hat{g}_0^R/(- \pi) +(\sigma^{R}_0)_{00}-(\sigma^{A}_0)_{00}}{(D^R + D^A)^2 + \left((\sigma^{R}_0)_{00}-(\sigma^{A}_0)_{00}\right)^2}, 
\end{equation} 
and $b(\varepsilon,T) = i v_f \frac{ -d T}{dx} \frac{ \varepsilon}{2 T^2} sech^2{\frac{\varepsilon}{2 T}}$. The second term is 
\begin{equation} \label{g1aKVSupp}
\hat{g}^{K,V}_{1a} (\hat{k},\varepsilon) = \hat{N}^R (\hat{g}^{R}_0 \hat{\sigma}^{K}_{1a}- \hat{\sigma}^{K}_{1a} \hat{g}^{A}_0).\end{equation}
$\hat{g}_{1a}^{K,V}$ is determined by $\hat{\sigma}_{1a}^K(\hat k) = n_i \hat{t}_{1a}^K(\hat k, \hat k)$. The strategy to obtain $\hat{t}^K_{1a}$ is to first ignore $\hat{g}^{K,V}_{1a}$ and calculate the non-self consistent $\hat{t}^{K,ns}_{1a}$ as well as $\hat{\sigma}^{K,ns}_{1a}$. Then, one can calculate $\hat{g}^{K,V}_{1a}$, which can be plugged back into the recurrence relation for $\hat t_{1a}^K$ in Eq.\ref{t1aKSupp} to obtain the self-consistency relations. The vertex correction is the only Gfnc to contribute to the thermal Hall conductivity, $\kappa_{yx}$, whereas $\hat{g}^{K,ns}_{1a}$ can only contribute to $\kappa_{xx}$ along the temperature gradient.

The energy current density is non-zero only for the anomalous Gfnc, $\hat{g}^{K}_{1a}$.
\begin{equation}\label{Jiy}
J_j = 2 N_f \int \frac{d\hat{k}}{4 \pi} v_{f,j} \int_{-\infty}^{\infty} \frac{d \varepsilon}{4 \pi i} \varepsilon \left( \hat{g}_{1a}^K \right)_{00}. 
\end{equation}

\section{Impurity Pair Breaking T-matrix approximation }\label{Supp2}
The first non-trivial finite size effect of a spherically symmetric impurity is the additional p-wave term on top of the s-wave scattering potential, $v(\hat{k},\hat{k}') = v_0 + v_1 \hat{k} \cdot \hat{k}'$. In spherical coordinates,
$$v(\hat{k},\hat{k}')= v_0 + 3 v_s \sin \theta \sin \theta' \cos (\phi - \phi') + 3 v_c \cos \theta \cos \theta'.$$
Note that, the imbalance between $v_1$ and $v_3$ allows for higher order terms scattering terms.

The impurities are included by the single impurity $t-matrix$ approximation with an impurity averaging approach. Each impurity independently and randomly modifies the system in the mean-field level. The impurities are invisible unless intentionally placed, therefore they cannot break the symmetries of the crystal but modify the system observables. It neglects the weak localization effect. The contribution from such impurity-impurity correlation is expected to be an order of magnitude smaller in the "small" expansion parameter, $(k_f l)^{-1}$, where $l$ is the typical impurity scattering length. 

The $t-matrix$ recurrence relation within the self-consistent full Born approximation is,
\begin{equation} \check{t} = v + N_f v \check{g} \check{t}. \end{equation}

Expanding the $t-matrix$ in the CoM gradients to the first order, $\check{t}(\hat{k},\hat{k}',\varepsilon) = \check{t}_0 (\hat{k},\hat{k}',\varepsilon) + \check{t}_1 (\hat{k},\hat{k}',\varepsilon)$, the recurrence relations for $RAK$ components are found as
\begin{eqnarray}
\label{tRZeroth}
\hat{t}^{R} &=& v + N_f \langle v \hat{g}^{R} \hat{t}^{R} \rangle_{\hat{k}''},\\
\label{tK1a1st}
\hat{t}^{K} &=& N_f \langle \hat{t}^{R} \hat{g}^{K} \hat{t}^{A} \rangle_{\hat{k}''} = (\hat{t}^{R} - \hat{t}^{A}) \tanh{\frac{\varepsilon}{2 T}} +\hat{t}^{K}_{1a}.\nonumber\\
\end{eqnarray}

In addition, the anomalous t-matrix is
\begin{equation}\label{t1aKSupp}
\hat{t}_{1a}^K = \pi N_f \langle \hat{t}_0^R \frac{\hat{g}_{1a}^K}{\pi} \hat{t}_0^A \rangle.
\end{equation}

Note that $\Delta(\hat k), \Gamma_u$ and $T/T_c$ are sufficient to calculate the THC coefficients, $\kappa_{ij}$.

\section{Thermal Conductivities} \label{Supp3}
\subsection{Normal State Limit}\label{Supp31}
The coefficients of the scattering potential in Eq.\ref{scatPot}, $v_0$, $v_s$ and $v_c$ are inputs to the transport equation. They are obtained from the normal state or Fermi liquid limit. Therefore, normal state limit has a fundamental role determining the inputs of the transport equation in superconducting limit. The physical observables are the phase shifts in a typical scattering experiment, the coefficients of the scattering potential, $v_0, v_s, v_c$ are parametrized in terms of the phase shifts of partial waves, $\cot{\delta_{0}} = \frac{-1}{ \pi N_f v_{0}}, \cot{\delta_{s}} = \frac{-1}{ \pi N_f v_{s}}, \cot{\delta_{c}} = \frac{-1}{ \pi N_f v_{c}}$

In the normal state limit, $\Delta_0 \to 0$, $RA$ components of the Gfncs are $\hat g_0^{R,A} = \mp i \pi \hat \tau_3$ and the impurity averages are 1, $\tilde{\gamma}_0 = \tilde{\gamma}_c = \tilde{\gamma}_s =1$. The t-matrix in Eq.\ref{tRZeroth} is easily determined,
\begin{widetext}
\begin{eqnarray*}
\hat t_{0,N}^{R} &=& \frac{-1}{\pi N_f}\left( \frac{\cot \delta_0 + i \hat \tau_3}{\cot^2 \delta_0 + 1}+ \frac{\cot \delta_s + i \hat \tau_3}{\cot^2 \delta_s + 1} \sin \theta \sin \theta' \cos(\phi - \phi') + \frac{\cot \delta_c + i \hat \tau_3}{\cot^2 \delta_c + 1} \cos \theta \cos \theta' \right) \\
&=& \frac{-1}{\pi N_f} \left(\sin \delta_0 e^{i \tau_3 \delta_0} +3 \sin \delta_s e^{i \tau_3 \delta_s} \sin \theta \sin \theta' \cos(\phi - \phi') +3 \sin \delta_c e^{i \tau_3 \delta_c} \cos \theta \cos \theta' \right)
\end{eqnarray*}

The most general zeroth order t-matrix in the normal state can be simplified into the partial waves with the corresponding phase shifts as
\begin{eqnarray*}
\hat t_{0,N}^{R} &=& \frac{-4 \pi}{\pi N_f} \sum_{l=0}^{\infty} \sum_{m=-l}^{l} \sin \delta_{lm} e^{i \delta_{lm} \hat \tau_3} Y^{m*}_l(\hat k') Y^{m}_l(\hat k) ,\\
&=& \frac{-1}{\pi N_f} \left(\sin \delta_{00} e^{i \delta_{00} \hat \tau_3} + 3 \sin \delta_{11} e^{i \delta_{11} \hat \tau_3} \sin \theta \sin \theta' \cos(\phi - \phi') + 3 \sin \delta_{10} e^{i \delta_{10} \hat \tau_3} \cos \theta \cos \theta' \right).
\end{eqnarray*}
Comparing each term (in our case $l_{max}=1$) and setting $\delta_{00} = \delta_{0}$ $\delta_{1,-1} = \delta_{11} = \delta_s$, $\delta_{10}=\delta_{c}$, it is verified that $\delta_i$s are actually the normal state phase shifts, which are also valid in superconductor state.

The normal state self-energy is $\hat \sigma_{0,N}^R = n_i \hat t_{0,N}^R$We can now also calculate the normal state Gfncs, $\hat{g}_{1a}^K$. Firstly, the non-self consistent part becomes,
\begin{equation*}
\hat g_{N,1a}^{K,ns} = \frac{\pi}{ \Im{\varepsilon^R}}\frac{ k_x}{k_f} b(\varepsilon,T).
\end{equation*}
Then, anomalous self-energy only has $\tau_0 \sigma_0$ component,
\begin{eqnarray}
\sigma_{N,1a}^K &=& \Gamma_u b(\varepsilon,T) 6 \sin \delta_0 \sin \delta_s \cos{(\delta_s-\delta_0)} \langle \frac{k_x^{''2}/k^2_f}{\Im{\varepsilon^R}} \rangle \frac{k_x}{k_f}
\end{eqnarray}
The vertex Gfnc is a scaled version of the non-self consistent Gfnc,
\begin{equation*}
g^{K,V}_{1a} \xrightarrow[\hat g_{N,1a}^{K} \to \hat g_{N,1a}^{K,ns}]{} \Gamma_u \left(6 \sin{\delta_0} \sin \delta_s \cos{(\delta_s-\delta_0)} \langle \frac{k_x^{''2}/k^2_f}{\Im{\varepsilon^R}}\rangle \right) \hat g_{N,1a}^{K,ns}
\end{equation*}
However, the full self-consistency demands a correction on the order of $\Gamma_u$. For this purpose let us define $Q_0 \equiv 6 \sin \delta_0 \sin \delta_s \cos{(\delta_s-\delta_0)}$ where $\hat g_{N,1a}^{K,V} = \Gamma_u Q \frac{\pi}{ \Im{\varepsilon^R}}\frac{ k_x}{k_f} b(\varepsilon,T)$ is the full self-consistent Gfnc. In the non-self-consistent limit, $Q \to Q_0 \equiv 6 \sin \delta_0 \sin \delta_s \cos{(\delta_s-\delta_0)}$ naturally. In this setting, if $\hat g_{N,1a}^{K,V}$ is plugged back into the Eq.\ref{t1aKSupp}, $Q$ is obtained self-consistently in terms of $Q_0$.
\begin{equation}
Q = \frac{Q_0}{1-\Gamma_u Q_0}.
\end{equation}
The thermal current is non-zero only for the longitudinal component, $K_{xx}$,
\begin{eqnarray}
\kappa^N_{xx} &=& \kappa_{xx}^{N,ns} +\kappa_{xx}^{N,V},\nonumber \\
\label{kxxN}
\frac{\kappa^N_{xx}}{\frac{\pi^2}{3}N_f v_f^2 T}&=& \frac{6}{\pi^2} \int d \tau \tau^2 sech^2{\tau} \Big[1 +\Gamma_u Q \langle \frac{k_x^{''2}/k^2_f}{\Im{\varepsilon^R}}\rangle \Big] \langle \frac{k_x^{''2}/k^2_f}{\Im{\varepsilon^R}}\rangle.
\end{eqnarray}
where $\tau = \varepsilon/2T$ is the dimensionless energy, $\Im{\varepsilon^R} = -\Im{(\sigma_0^R)_{30}} = - \gamma(\theta)$ where $\gamma(\theta) = - \gamma_0 - \gamma_s \sin^2 \theta - \gamma_c \cos^2 \theta = \Gamma_u \left( \sin^2 \delta_0+ 3 \sin^2 \delta_s \sin^2 \theta+ 3 \sin^2 \delta_c \cos^2 \theta \right)$. The $\Gamma_u$ dependent expression is obtained by the self-consistency relations for $\gamma_0,\gamma_s$ and $\gamma_c$ in Eq.s \ref{selfConsE1g} and \ref{selfConsE2g}.

In Fig.\ref{KxxKxxNVsTE1gE2g}, we plot the $\kappa_{xx}/\kappa^N$ as a function of temperature. At $T/T_c =1$, $\kappa_{xx}/\kappa^N \to 1$, where the exact normal state limits are reproduced both for $E_{1g}$ and $E_{2g}$ as given in Eq.\ref{kxxN}. Keep in mind that either $\delta_c$ or $\delta_s$ are omitted in our work, then the integral averages multiplied with $\Gamma_u$ are evaluated as
\begin{eqnarray}
E_{1g} (\delta_s \to 0):&& \quad \Gamma_u \langle \frac{k_x^{2}/k^2_f}{\Im{\varepsilon^R}}\rangle = \frac{1}{2}\left[ \left( 1+\frac{1}{c} \right) \frac{\arctan \sqrt{\frac{c}{a}}}{\sqrt{ac}} - \frac{1}{c} \right] \equiv \langle..\rangle_{E_{1g}},\\
E_{2g} (\delta_c \to 0):&& \quad \Gamma_u \langle \frac{k_x^{2}/k^2_f}{\Im{\varepsilon^R}}\rangle = \frac{1}{2}\left[ \left( 1-\frac{1}{b} \right) \frac{arctanh \sqrt{\frac{b}{a+b}}}{\sqrt{(a+b) b}} + \frac{1}{b} \right] \equiv \langle..\rangle_{E_{2g}}
\end{eqnarray}
where $(a,b,c) \equiv (\sin^2 \delta_0,3\sin^2 \delta_s,3\sin^2 \delta_c)$. The averages, $\langle..\rangle_{E_{1g}}$ and $\langle..\rangle_{E_{2g}}$ hence, are independent of the impurity concentration. Finally, normalizing both side of Eq.\ref{kxxN} by $\Gamma_u$, we can obtain $\kappa^N_{xx}$ for $E_{1g}$ and $E_{2g}$ gap parameters, 
\begin{eqnarray}
\label{KxxNE1g}
\frac{\kappa^{N,E_{1g}}_{xx} \Gamma_u }{ \frac{\pi^2}{3}N_f v_f^2 T} &=& \frac{6}{\pi^2} \int d \tau \tau^2 sech^2{\tau} \langle..\rangle_{E_{1g}},\\
\label{KxxNE2g}
\frac{\kappa^{N,E_{2g}}_{xx} \Gamma_u }{\frac{\pi^2}{3}N_f v_f^2 T} &=& \frac{6}{\pi^2} \int d \tau \tau^2 sech^2{\tau} \left( 1 + \frac{Q_0}{1 - \Gamma_u Q_0}\langle..\rangle_{E_{2g}} \right) \langle..\rangle_{E_{2g}}.
\end{eqnarray}
The two final expressions are useful not only to discuss the longitudinal currents and the superconducting contribution but also to analyse the thermal Hall conductivity as a normalization factor when the impurity concentration, $\Gamma_u$, dependence of $\kappa_{yx}$ is discussed in the last section of the Appendices. 

\subsection{THCs for the order parameter $\Delta^{E_1g}_k = \Delta_0 \cos{\theta} \sin{\theta} e^{i \phi}$}\label{Supp32}
\subsubsection{$E_{1g}$, the zeroth order t-matrix and the self-energy}\label{Supp321}
Expanding the Born series and collecting the common terms for $E_{1g}$, the general form of the $\hat t_0^R$ matrix can be found by inspection as,
\begin{eqnarray}\label{t0RE1g}
\hat{t}_0^R (\hat{k},\hat{k}') &= & \hat{t}^{R}_{00} + \left[\hat{t}^{R}_{ssc} \cos{(\phi-\phi')} + \hat{t}^{R}_{+} e^{i (\phi -\phi')\tau_3} \right] \sin \theta \sin \theta'+\hat{t}^{R}_{cc} \cos \theta \cos\theta' \nonumber \\
&+& \left( \hat A_1 \cos \theta \sin \theta' e^{i \phi' \tau_3} + \hat A_2 \cos \theta' \sin \theta e^{i \phi \tau_3} \right) \hat \tau_1 \hat \sigma_2 .
\end{eqnarray}
\begin{figure*}
\includegraphics[width=0.97\textwidth]{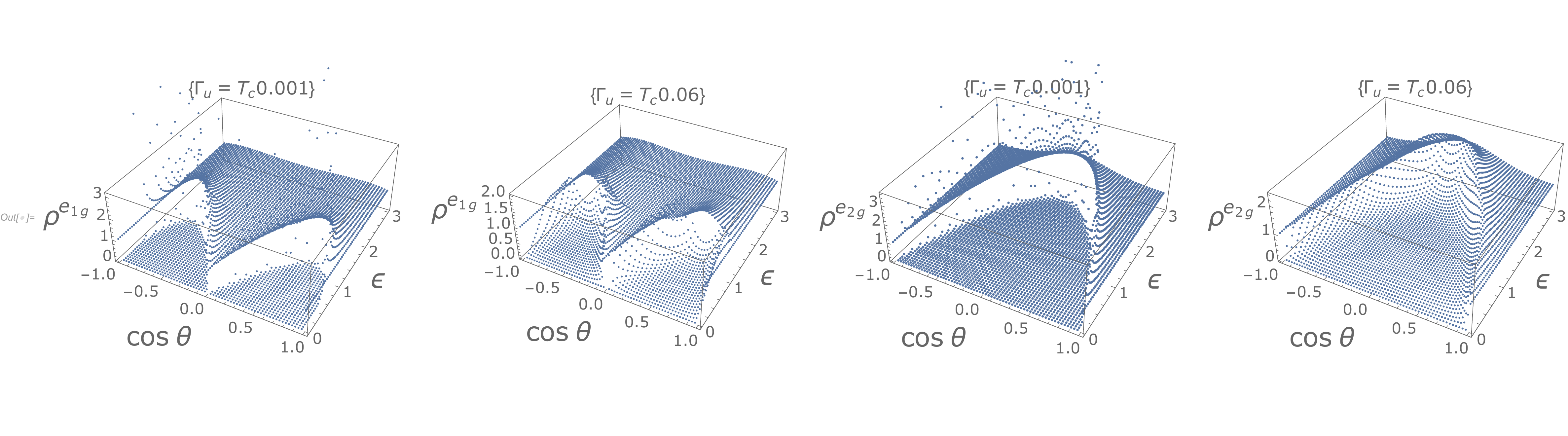}
\caption{Density of states for $\Delta^{E_{1g}}$ and $\Delta^{E_{2g}}$ gap functions in the space of $\{ \cos \theta - \varepsilon \}$, where $x$ is the polar angle. First two plots on the left are for $\Delta^{E_{1g}}$, the density of states for the clean limit ($\Gamma_u = 0.001$ T$_c$) and the finite impurity ($\Gamma_u = 0.06$ T$_c$), where  $T=0.4$ T$_c$, and the two right most are for $\Delta^{E_{2g}}$. The phase shifts are equal, $\delta_0 = \delta_s = \delta_c = \pi/3$.}
\label{ModifiedDOS}
\end{figure*}

Plugging $\hat t_0^R(\hat k,\hat k')$ in Eq.\ref{t0RE1g} back into the recurrence relation and matching the terms with respect to the angular functions and Nambu space components $\hat \tau_i \hat \sigma_j$, we obtain the elements of the $\hat t_0^R(\hat k,\hat k')$ as
\begin{equation*}
\begin{split}
\hat{t}^{R}_{00} &= \frac{-1}{\pi N_f} \frac{\cot \delta_0 + i \tilde \gamma_{0} \hat \tau_3}{C_{0}},\quad \hat{t}^{R}_{ssc} = \frac{-3}{\pi N_f} \frac{\cot \delta_s + i \tilde \gamma_{s} \hat \tau_3}{C_s} \\
\hat{t}^{R}_{cc} &= \frac{-3}{\pi N_f}\left[\cot \delta_c - i \tilde{\gamma}_{c} \tau_3+ \frac{2 (\tilde{\gamma}^1)^2}{\cot \delta_s + i \tilde{\gamma}_s \tau_3} \right]^{-1}\\
\hat{t}^{R}_{+} &= \frac{-3}{\pi N_f} \frac{ \tilde\gamma^1 \cot \delta_c}{(\cot \delta_c + i \tilde{\gamma}_c \tau_3) (\cot \delta_s - i \tilde{\gamma}_s \tau_3) \cot \delta_c \cot \delta_s + 2 (\tilde{\gamma}^1)^2} \\
\hat A_{1} &= - (\cot \delta_s +i \tilde{\gamma}_s \tau_3) \hat{t}^{R,-}_{+},\quad \hat A_{2} = \frac{ \pi N_f \tilde\gamma^1}{3}\hat{t}^{R}_{ssc} \hat{t}^{R}_{cc} 
\end{split}
\xrightarrow[where]{}
\begin{split}
C_{0} &= \cot^2 \delta_0 + (\tilde\gamma_{0})^2, \\
C_{s} &= \cot^2 \delta_s + (\tilde\gamma_{s})^2, \\
C_{c} &= \cot^2 \delta_c + (\tilde\gamma_{c})^2,\\
\hat{t}^{R,-}_{+} & \equiv (\hat{t}^{R}_+)_{00} - (\hat{t}^{R}_+)_{30} \tau_3.
\end{split}
\end{equation*}

The integrals are different momentum averages of the Gfncs are,
\begin{eqnarray}\label{selfConsis}
\tilde \gamma_{0} = \langle \frac{- i\varepsilon^R(\theta) }{D^R(\theta)} \rangle&,&\quad \tilde \gamma^1 = \frac{3}{2} \langle \frac{\Delta_0 \sin^2 \theta \cos^2 \theta}{D^R(\theta)} \rangle,\nonumber\\\\
\tilde \gamma_{c} = 3 \langle \frac{- i\varepsilon^R(\theta) }{D^R(\theta)} \cos^2 \theta \rangle&,&\quad \tilde \gamma_{s} = \frac{3}{2}\langle \frac{- i\varepsilon^R(\theta) }{D^R(\theta)} \sin^2 \theta \rangle. \nonumber\\
\end{eqnarray}
\end{widetext}
Then, the zeroth order self-energy is,
\begin{eqnarray}
\hat \sigma^R_0 (\hat k)&=& n_i \hat t^R (\hat k, \hat k),\nonumber\\
\label{sigma0RE1g}
&=& \pi N_f \Gamma_u \Big[ \hat{t}^{R}_{00} + \left( \hat{t}^{R}_{ssc} + \hat{t}^{R}_{+} \right) \sin^2 \theta + \hat{t}^{R}_{cc} \cos^2 \theta \nonumber + \cos \theta \sin \theta (\hat A_1 + \hat A_2 ) e^{i \phi \tau_3}\tau_1 i \hat \sigma_2\Big].
\end{eqnarray}
where it is an even function of $\hat k$. Using $\hat \sigma^R_0 (\hat k)$ in Eq.\ref{sigma0RE1g}, the self-consistency relation for the diagonal element is found to be, 
\begin{eqnarray}\label{selfConsE1g}
\varepsilon^R = \varepsilon - \pi N_f \Gamma_u \Big( (\hat{t}^{R}_{00})_{30}&+&\left[ (\hat{t}^{R}_{ssc})_{30} +(\hat{t}^{R}_{+})_{30}\right] \sin^2 \theta \nonumber \\ + (\hat{t}^{R}_{cc})_{30} \cos^2 \theta  \Big).
\end{eqnarray}
Another relation is for the off-diagonal terms, $\Delta_0^R = \Delta_0 + \pi N_f \Gamma_u (A_1 + A_2 )$. 

Inspecting the t-matrix, it should be emphasized that $\delta_c$ plays the most important role. Along with $\delta_s$, they lead to a spontaneous skew scattering \citep{skewScatteringAHE} channel, $\hat{t}^{R}_{+}$ which contributes both diagonal and off diagonal part of the t-matrix. If it goes to zero, t-matrix is reduced to two channels $\hat{t}^{R}_{00},\hat{t}^{R}_{ssc}$ which is inadequate to produce a finite Hall current. Actually, it is possible to retain the finite $\kappa_{yx}$ only with $\delta_c$ component as will be shown at the beginning of next subsection. Even if $\delta_s$ is introduced back, it only renormalizes the contributions due to $\delta_c$.

The density of states $\rho(\cos{\theta},\varepsilon)$ can be calculated in the presence of impurities. The (un)modified distribution is shown in Fig.\ref{ModifiedDOS} for equal phase shifts, $\delta_0 = \delta_1 = \delta_3 = \pi/3$ and the temperature is $T = 0.4$ T$_c$. The first plot is the clean limit DOS while the second plot is for the impurity concentration $\Gamma_u = 0.06$ T$_c$. The modified DOS graphs shows the new BQs are due to the broken pairs on the Fermi level. The equator line and the two polar nodes are modified with finite DOS even at low temperatures, which clearly indicates the formation of the impurity band as a new energy scale as $\varepsilon \to 0$. Hence, there are BQs available for thermal transport even at low temperatures.

\subsubsection{$E_{1g}$ Anomalous t-matrix and the self-energy}\label{Supp322}
Using Eq.\ref{t1aK}, the anomalous t-matrix $\hat{t}^{K,ns}_{1a}$ can be calculated with a known initial condition, which is $\hat{g}^{K,ns}_{1a}$ given in Eq.\ref{g1aKnsSupp}.It vanishes in the absence of the external field, the temperature gradient. The whole procedure including the self-consistent equation to calculate the vertex correction Gfnc is straightforward. In the first step, $g_{1a}^{K} \to g_{1a}^{K,ns}$, as the initial condition. It is called the non-self-consistent solution, corresponding to $t_{1a}^{K,ns}$. In the second step, plug $g_{1a}^{K} = g_{1a}^{K,V} $.Then, the resulting $t-matrix$ is $\hat{\bar{t}}_{1a}^{K}=\hat{t}_{1a}^{K}$ and $\hat{t}_{1a}^{K,ns}$. Comparing the the coefficients of the self-energies $\hat{\sigma}_{1a}^{K},\hat{\bar{ \sigma}}_{1a}^{K} $, the self-consistency relations can be solved for the full-self consistent result for $\hat{g}_{1a}^K$ as $\hat{\bar{\sigma}}_{1a}^{K} =\hat{\sigma}_{1a}^{K}-\hat{\sigma}_{1a}^{K,ns} = n_i N_f \langle \hat t_0^R \hat g_{1a}^{K,V} \hat t_0^A \rangle $.

The scattering potential, $v(\hat k,\hat k')$ is effectively present for quasiparticle momentum states around the equator of the Fermi level, where the gap function $\Delta(\hat k)$ is suppressed with a line node. Considering the effect of only the anisotropic term $v_s$, the gapless nodes with momentum $\hat{k}' \sim (\cos \phi' \hat x + \sin \phi' \hat y)$ are strongly scattered around the same horizontal plane dividing the equator at $\theta \sim \pi/2$. The effective scattering potential is 2-d,
$v(\hat k,\hat k') \approx v_0 + 3 v_s \cos (\phi-\phi')$ and has a rotational symmetry. In this limit, however, the Hall conductivity is zero, which can be understood as follows. The anomalous self energy for equator scattering case is,
\begin{eqnarray*}
\hat{\sigma}^K_{1a} &\sim& \langle \hat{t}_0^R \hat{g}_{1a}^{K,ns} \hat{t}_0^A \rangle\\
&\sim& \left(\hat{t}_0^R\right)_{even} \langle \left(\hat{g}_{1a}^{K,ns}\right)_{diag} \left(\hat{t}_0^A\right)_{odd} \rangle \frac{k_x}{k_f}\\
&&+ \langle \left(\hat{t}_0^R\right)_{odd} \left(\hat{g}_{1a}^{K,ns}\right)_{diag} \rangle \left(\hat{t}_0^A\right)_{even}\frac{k_x}{k_f} \\
&=& \Gamma_u (\hat{...})_{even} \sin \theta \cos{\phi} b(\varepsilon,T).
\end{eqnarray*} 
The even and odd parts of the t-matrix are $(\hat t_0^{R,A})_{even} = \hat t_{00}^{R,A}$ and $(\hat t_0^{R,A})_{odd} = \hat t_0^{R,A}- \hat t_{00}^{R,A}$. Using Eq.\ref{g1aKVSupp} the anomalous Gfnc has the form,
\begin{eqnarray*}
\hat{g}^{K,V}_{1a} &=& \Gamma_u \hat{N}_R \left( \hat{g}_0^R (\hat{...})_{even} - (\hat{...})_{even} \hat{g}_{0}^{A} \right)\frac{k_x}{k_f} b(\varepsilon,T),\\
&\sim& \Gamma_u (\hat{..})_{even}\frac{k_x}{k_f} b(\varepsilon,T).
\end{eqnarray*} 
As seen clearly, all components of $\hat{g}^{K,V}_{1a}$ is proportional to $k_x$ whereas the non-zero Hall current density strictly requires $v_{f,y}\sim k_y$ component. Therefore, $J_y \sim \int d \hat k v_{f,y} k_x (..) = 0$.

\textbf{Below, we therefore consider only $v_0,v_c$}. The polar scattering potential $v(k,k') = v_0 + 3 v_c \cos \theta \cos \theta '$ is considered as the dominant scattering process. It gives rise to finite $\kappa_{yx}$. It preserves the rotational symmetry of the system and does not break any extra symmetry even when the equator scattering part is dropped. The t-matrix becomes,
\begin{widetext}
\begin{equation}
\hat t^R_0(\hat k, \hat k')= \hat t_{00}^R + \hat t_{cc}^R \cos \theta \cos \theta'= \frac{-1}{\pi N_f} \Big( \frac{\cot \delta_0 +i \tilde \gamma_{0} \hat \tau_3}{C_{0}}.
+ 3 \frac{\cot \delta_c + i \tilde \gamma_{c} \hat \tau_3}{C_{c}} \cos \theta \cos \theta'\Big).
\end{equation}
Note that $\hat{t}^A(\hat k, \hat k') = \hat \tau_3 \left(\hat{t}^R(\hat k', \hat k) \right)^\dagger \hat \tau_3$.
- Step 1:
The non-self consistent anomalous Gfnc, $g_{1a}^{K,ns}$ has the following form, 
\begin{equation}\label{g1aknsExplicit}
\hat{g}_{1a}^{K,ns} = \pi \left( \frac{2 \Re{D^R}}{D_1} \left[ \left(1+\frac{\lvert \varepsilon^R \rvert^2 - \lvert \Delta(\hat k)\rvert^2}{\lvert D^R \rvert^2} \right)-2 i \frac{\Im{\varepsilon^R} \hat \tau_3\hat{\Delta}}{\lvert D^R \rvert^2} \right]+\frac{4 \Im{\sigma_{00}^R}}{D_1} \left[ \Im{\frac{\varepsilon^R}{D^R}} \hat \tau_3-\Im{\frac{1}{D^R}}\hat{\Delta} \right]\right) \sin \theta \cos \phi b(\varepsilon,T).
\end{equation}
Plugging $g_{1a}^{K,ns}$ into the t-matrix relation in Eq.\ref{t1aK}, the non-vanishing terms arise from the off-diagonal part of $\hat{g}_{1a}^{K,ns}$ as the integrals have the form $\sim \langle \hat{g}_{1a}^{K,ns} \frac{k_z}{k_f} \rangle$ which is clarified below. Taking $\hat{\sigma}_{1a,ns}^K=n_i \hat{t}_{1a}^K(\hat k, \hat k)$ limit of the anomalous t-matrix, $\hat{\sigma}_{1a,ns}^K$ is a odd function of $\hat k$ and proportional to $ \frac{k_z}{k_f}$,
\begin{equation}
\hat\sigma_{1a}^{K,ns}(\hat k) = \Gamma_u \hat t^{-}_{0c} \left( \beta_0 + i \beta_3 \hat \tau_3 \right) \hat \tau_1 i \hat \sigma_2 \frac{k_z}{k_f} b(\varepsilon,T),
\end{equation}
where $\hat{t}^{-}_{0c} \equiv (\pi N_f)^2 \left( \hat{t}^{R}_{00} \hat{t}^{A-}_{cc}+\hat{t}^{R}_{cc}\hat{t}^{A-}_{00} \right)/2 = (\hat t^{-}_{0c})_{00} + (\hat t^{-}_{0c})_{30} \hat \tau_3$. The averages are the various integrals of the $\hat \tau_i \hat \sigma_j$ components of the non-self consistent anomalous Gfns,
\begin{eqnarray}\label{integralBeta}
\beta_0 (\varepsilon) &=& \langle \frac{4\Im{ \sigma_{00}^R}}{D_1} \Im{\frac{\Delta_0}{D^R}} \cos^2 \theta \sin^2 \theta \rangle,\\
\beta_3 (\varepsilon) &=& \langle \frac{4 \Re{D^R}}{D_1}\frac{\Im{\varepsilon^R} \Delta_0 }{\lvert D^R \rvert^2 }\cos^2 \theta \sin^2 \theta \rangle.
\end{eqnarray}
Note that $D_1 = 4(\Re{D^R})^2 + 4\Im{(\sigma^{R}_0)_{00}}^2$ and $\hat t_{00}^{A,\pm} = (\hat t_{00}^A)_{00} \pm (\hat t_{00}^A)_{30} \tau_3$. 

We simplify $\hat{\sigma}_{1a}^K$ by defining,
\begin{equation}
\begin{bmatrix}
G_0 \\ G_3
\end{bmatrix} =\begin{bmatrix}
(\hat{t}^{-}_{0c})_{00} \beta_0 + (i \hat{t}^{-}_{0c})_{30} \beta_3 \\(\hat{t}^{-}_{0c})_{00} \beta_3- ( i\hat{t}^{-}_{0c})_{30} \beta_0.
\end{bmatrix} 
\end{equation}
The self energy becomes,
\begin{eqnarray}
\hat\sigma_{1a}^{K,ns}(\hat k) &=& \Gamma_u \hat G \hat \tau_1 i \hat \sigma_2 \frac{k_z}{k_f} b(\varepsilon,T),
\end{eqnarray}
where $\hat G = G_0 + i G_3 \hat \tau_3 $. Plugging $\hat{\sigma}_{1a}^K$ into Eq.\ref{g1aKVSupp}, $\hat g_{1a}^{K,V}$ is obtained in terms of $\hat G$. We know that $\hat G$ should be replaced by $\hat{\tilde{G}}$ in the full self-consistent case. Its explicit form is long and will not be given here in detail, but it can be cast into the following form,
\begin{eqnarray}\label{g1aKVSuppE1g}
\hat{g}^{K,V}_{1a} (\hat{k},\varepsilon) &=& \pi \Gamma_u b(\varepsilon, T) \sin(\theta) \left[ \hat X_c \cos \phi + \hat X_s \sin \phi + \hat X_{2\phi} e^{i 2 \phi \tau_3} + \hat Y e^{i \phi \hat \tau_3} \tau_1 i\sigma_2 \right].
\end{eqnarray}
Note that all $\hat X_i$s are diagonal matrices, which can be written as linear combination of $G_0$ and $G_3$. In addition, the vertex corrections modify these coefficients, let us denote them as $\tilde{G}_0,\tilde{G}_3$, and consequently the coefficients of the vertex correction Gfnc are also modified.
\begin{figure}
\includegraphics[width=0.50
\textwidth]{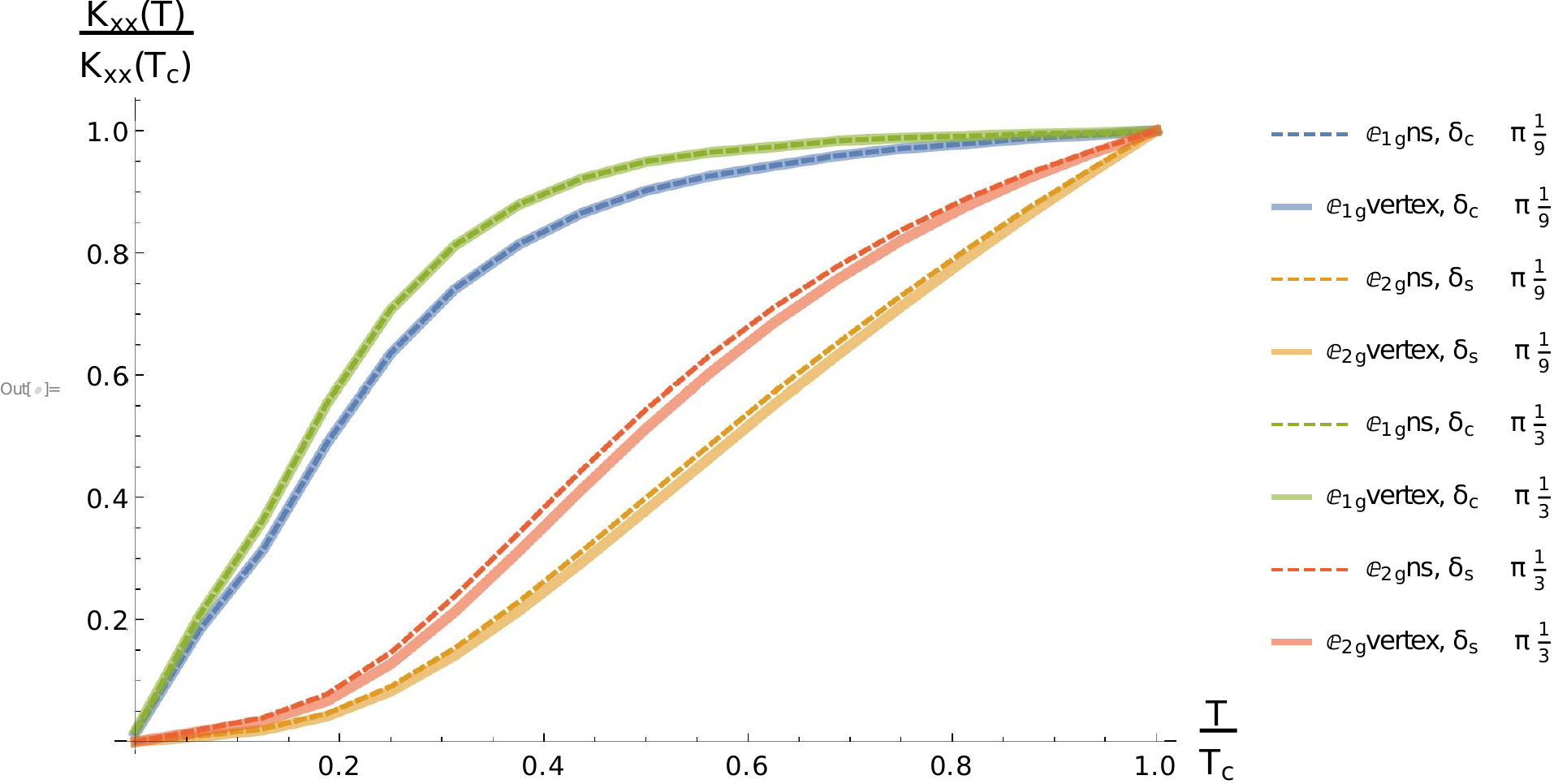}
\caption{Longitudinal thermal conductance, $\kappa_{xx}$ as a function of temperature (in units of $T_c$), for the non-self consistent part $\kappa^{ns}_{xx}(T)/\kappa^{ns}_{xx}(T_c)$ (dashed lines) only and for $\kappa_{xx}(T)/\kappa_{xx}(T_c)$ (solid lines) where the vertex correction is included. The upper two pairs of lines are for $E_{1g}$ and lower two pairs are for $E_{2g}$. For both cases $\delta_0 = \pi/3$ and $\delta_s \in \{\pi/9,\pi/3\}$. The unity limit is reproduced at $T=T_c$. The vertex correction is significant only for non-zero $\delta_s$, in this case for $E_{2g}$.}
\label{KxxKxxNVsTE1gE2g}
\end{figure}

- Step 2:
The same procedure can be repeated if we replace $\hat g_{1a}^{K} \to \hat g_{1a}^{K,V}$. The corresponding t-matrix is $\hat{\bar{t}}_{1a}^{K} = \hat{t}_{1a}^{K}-\hat{t}_{1a}^{K,ns}$. The only non-vanishing averages in the renormalization comes from the off-diagonal components of $\hat{g}_{1a}^K$ in Eq.\ref{g1aKVSuppE1g}, $\hat Y$ which is given as 
\begin{equation}\label{YcoeffE1g}
\hat Y = -4 \left( \frac{\Re{D^R}}{D_1}(1- \lvert \frac{\varepsilon^R}{D^R} \rvert^2) - 2 i \frac{\Im{\sigma_{00}^R}}{D_1}\Re{\frac{\varepsilon^R}{D^R} \tau_3} \right)\hat{\tilde{G}}.
\end{equation}
In this way, the anomalous self-energy can be obtained with the following relation $\bar{z}_i = \tilde{z}_i - z_i$,
\begin{equation}\label{vertexCorrSelfE1g}
\begin{bmatrix}
\tilde{G}_0 \\ \tilde{G}_3
\end{bmatrix} = \frac{1}{Det} \begin{bmatrix}
1 + \Gamma_u F_0 & \Gamma_u F_1 \\-\Gamma_u F_1 & 1 + \Gamma_u F_0
\end{bmatrix} \begin{bmatrix}
G_0 \\ G_3
\end{bmatrix}.
\end{equation}
where $Det = 1 + 2 \Gamma_u F_0 + \Gamma^2_u (F_0^2 + F_1^2)$.

\begin{equation}
\begin{bmatrix}
F_0 \\ F_1
\end{bmatrix} = \begin{bmatrix}
\lambda_0 (t^{-}_{0c})_{00} + \lambda_3 (t^{-}_{0c})_{30} \\\lambda_3 (t^{-}_{0c})_{00} - \lambda_0 (t^{-}_{0c})_{30}
\end{bmatrix} \end{equation}
where the integrals are defined as,
\begin{equation}
\begin{bmatrix}
\lambda_0(\varepsilon) \\ \lambda_3(\varepsilon)
\end{bmatrix} = 4 \langle \frac{1}{D_1} \begin{bmatrix}
\Re{D^R}(1-\lvert \varepsilon^R/D^R\rvert^2) \\ 2\Im{\sigma^R_{00}}\Re{(\varepsilon^R/D^R)})
\end{bmatrix}\cos^2 \theta\rangle.
\end{equation}
We finally obtain all necessary terms to calculate the components of current density. The current density, $J_i$ is proportional to $\tau_0 \sigma_0$ component of $\hat g_{1a}^{K}$. The non-self consistent part, $\hat g_{1a}^{K,ns}$ yields only non-zero longitudinal conductivity,
\begin{eqnarray}\label{JxnsE1gSupp}
\frac{\kappa^{ns}_{xx} T_c}{\frac{\pi^2}{3} N_f v_f^2 T} &=&T_c \frac{6 }{\pi^2} \int_{-\infty}^{\infty} \frac{d \varepsilon}{2 T}\frac{\varepsilon^2}{4 T^2} sech^2(\frac{\varepsilon}{2 T}) \Omega_0(\varepsilon).
\end{eqnarray}
The integral average is defined as $\Omega_0(\varepsilon) = \langle \frac{2 \Re{D^R}}{D_1} \left(1+ \frac{ \lvert \varepsilon^R \rvert^2 - \lvert \Delta \rvert^2}{\lvert D^R \rvert^2} \right) \sin^2 \theta \rangle_{\theta}$. The whole expression is dimensionless. 

Omitting $2 \phi$ dependent terms of $\tau_0$ component as the corresponding integrals would vanish, $(\hat{g}_{1a}^{K,V})_{00}$ becomes, 
\begin{eqnarray}
(\hat g_{1a}^{K,V})_{00} &=& \pi \Gamma_u b(\varepsilon,T) \left[ X_{c,0} \frac{k_x}{k_f} + X_{s,0} \frac{k_y}{k_f} \right],\\
\begin{bmatrix}
X_{c,0} \\ X_{s,0}
\end{bmatrix} &=& \frac{-8 \cos^2 \theta \Delta_0 }{D_1}\begin{bmatrix}
\Im{\sigma_{00}^R} \Im{1/D^R}& \frac{-\Re{D^R}\Im{\varepsilon^R}}{\lvert D^R\rvert^2} \\ \frac{\Re{D^R}\Im{\varepsilon^R}}{\lvert D^R\rvert^2} & \Im{\sigma_{00}^R}\Im{1/D^R} 
\end{bmatrix} \begin{bmatrix}
\tilde{G}_{0} \\ \tilde{G}_{3}
\end{bmatrix}.
\end{eqnarray}
Then, the conductivity elements are found to be,
\begin{eqnarray}
\frac{\begin{bmatrix}\kappa^V_{xx}\\ \kappa^V_{yx}\end{bmatrix}}{\frac{\pi^2}{3} N_f v_f^2 T} &=& \Gamma_u \frac{6}{\pi^2} \int_{-\infty}^{\infty}\frac{ d \varepsilon}{2 T} \left(\frac{\varepsilon}{2 T} \right)^2 sech^2(\frac{\varepsilon}{2 T}) \frac{1}{4} \langle \sin^2 \theta \begin{bmatrix}
X_{c,0}(\varepsilon,\theta) \\ X_{s,0}(\varepsilon,\theta)
\end{bmatrix} \rangle\\
\label{KyxKxxE2gSupp}
&=& \Gamma_u \frac{6}{\pi^2} \int_{-\infty}^{\infty} d \tau \tau^2 sech^2(\tau) (-2) \begin{bmatrix}
\beta_0(2 \tau T) \tilde{G}_0 - \beta_3(2 \tau T) \tilde{G}_3 \\ \beta_3(2 \tau T) \tilde{G}_0 + \beta_0(2 \tau T) \tilde{G}_3
\end{bmatrix} \xrightarrow[\varepsilon,T \to 0]{} \Gamma_u \beta_3(0) 2 \begin{bmatrix}\tilde{G}_3\\ - \tilde{G}_0\end{bmatrix}.
\end{eqnarray}

In the low temperature limit, $\varepsilon^{R,A} \to i \gamma(\theta)$, $D^{R,A} = D \to \sqrt{\lvert \Delta(\hat k)\rvert^2 + \gamma^2(\theta)}$, and the following two integrals vanish $\beta_0(0), \lambda_3(0) \to 0$. Also, $\Omega_0(0)\to \langle \frac{\gamma^2}{ D^3} \sin^2 \theta \rangle \sim \frac{\gamma_0}{\Delta^2_0}$, $\beta_3(0) \to \langle \frac{\gamma \Delta_0}{ D^2} \cos^2 \theta \sin^2 \theta \rangle \sim \gamma_0/\Delta^2_0$, $\lambda_0(0) = \langle \frac{\lvert \Delta(\hat k) \rvert^2}{D^3} \cos^2 \theta \rangle \sim \frac{1}{\Delta_0} \ln \frac{\Delta_0}{\gamma_0}$, and $\{ F_0, F_1 \} \to \lambda_0 \{(t^{-}_{0c})_{00}, -(t^{-}_{0c})_{30} \}$, $Det \to 1 + \Gamma_u 2 \lambda_0 (t^{-}_{0c})_{00} + \Gamma_u^2 \lambda_0^2 \left[ ((t^{-}_{0c})_{00})^2+((t^{-}_{0c})_{30})^2\right] \approx 1 + 4 (t^{-}_{0c})_{00} \frac{\Gamma_u}{\Delta_0} \ln \frac{\Delta_0}{\gamma_0}$.
$$ \begin{bmatrix} (t^{-}_{0c})_{00}\\(t^{-}_{0c})_{30}^{-}\end{bmatrix} \to \frac{1}{C_0 C_c} \begin{bmatrix}\cot \delta_0 \cot\delta_c - \tilde{\gamma}_{0}\tilde{\gamma}_{c}\\\cot \delta_0 \tilde{\gamma}_{c} +\cot\delta_c \tilde{\gamma}_{0}\end{bmatrix}$$

The conductivities are found to be,
\begin{eqnarray}
\frac{\begin{bmatrix}
\kappa_{xx}\\\kappa_{yx}\end{bmatrix}T_c}{\frac{ \pi^2 }{3}N_f v^2_f T} &=& \begin{bmatrix}T_c\Omega_0(0)
\\0\end{bmatrix}+ \Gamma_u T_c \beta_3(0) 2 \begin{bmatrix}\tilde{z}_3\\ - \tilde{z}_0\end{bmatrix},\\
\label{KyxE1gLowTempSupp}
&\approx& \begin{bmatrix} T_c \Omega_0(0)
\\0\end{bmatrix}+ \Gamma_u \frac{ T_c}{C_0 C_c}\begin{bmatrix}\cot \delta_0 \cot\delta_c - \tilde{\gamma}_{0}\tilde{\gamma}_{c}\\ \cot \delta_0 \tilde{\gamma}_{c}+\cot\delta_c \tilde{\gamma}_{0}\end{bmatrix}\frac{4 \beta_3^2(0)}{Det},\\
\label{KyxE1gLowTempApproxSupp}
& \sim& \frac{T_c}{\Delta_0} \frac{\gamma_0^2 }{\Delta^2_0}  \ln{\frac{\Delta_0}{\gamma_0}} \begin{bmatrix} \cot \delta_0 \cot \delta_c -  \frac{\gamma^2_0}{\Delta^2_0} \ln{\frac{\Delta_0}{\gamma_0}}\\ \frac{\gamma_0}{\Delta_0}\left( \cot \delta_0 +  \cot \delta_c \ln{\frac{\Delta_0}{\gamma_0}}\right) \end{bmatrix}\frac{1}{C_c}.
\end{eqnarray}
Note that Eq.\ref{KyxE1gLowTempApproxSupp} are the upper limits for $\kappa_{ij}$ obtained for $\cot \delta_i \sim \tilde{\gamma}_i$, $i \in \{ 0,c \}$ along with $Det \sim 1$ and the self-consistency relations in Eq.\ref{selfConsis}.

To get more insight into the expressions for $\kappa_{ij}$ in Eq.\ref{KyxE1gLowTempSupp}, but if $\theta$ is fixed, $\theta = \theta_0$ and $\theta = \pi- \theta_0$, the order parameter becomes identical to a p-wave superconductor with s-wave scattering. The scattering potential takes a constant value, $V = v_0 + 3 v_c \cos^2 \theta_0$ in the upper hemisphere (or lower hemisphere), while it does not scatter between $\theta_0$ and $\pi-\theta_0$ if we choose $v_0 = 3 v_c \cos^2 \theta_0$. Then the vertex correction to the conductivities converge to the result, \citep{yip2016low4} as
\begin{eqnarray}
\frac{\begin{bmatrix}\kappa^V_{xx}\\\kappa^V_{yx}\end{bmatrix}}{\frac{ \pi^2 }{3}N_f T} &\propto& \frac{\Gamma_u}{Det} \begin{bmatrix}\cot^2 \delta_0 - (\tilde{\gamma}_{0})^2 \\ 2 \cot \delta_0 \tilde{\gamma}_{0}\end{bmatrix}\frac{\gamma^2_0}{ C^2_{0}}v^2_f \langle \frac{\Delta_0 \lvert f(\hat k)\rvert^2}{D^3} \rangle^2,
\end{eqnarray}
where the momentum dependent part of the order parameter is separated as $\Delta(\hat k) = \Delta_0 f(\hat k)$ and $f(\hat k)$ is the representation in k-space. For simplicity, we assumed the impurity band to be constant, $\gamma(\theta) \approx \gamma_0$.

\subsection{THCs for the order parameter $\Delta^{E_2g}_k = \Delta_0 \sin^2{\theta} e^{i 2 \phi}$}\label{Supp33}
\subsubsection{$E_{2g}$, The zeroth order t-matrix and the self-energy}\label{Supp331}
Expanding the Born series for $\Delta^{E_{2g}}_k$ gap symmetry, and by inspection, the zeroth order t-matrix takes the form,
\begin{eqnarray*}
\hat{t}_0^R &=& \hat{t}^{R}_{00} + \hat{t}^{R}_{ssc} \sin \theta \sin \theta' \left( \cos{(\phi-\phi')}- \frac{\alpha}{2} e^{i (\phi-\phi')\hat \tau_3}\right) + \hat{t}^{R}_{cc} \cos \theta \cos \theta'- \frac{3\alpha_0}{2 \pi N_f} e^{i (\phi + \phi') \hat \tau_3}\hat \tau_1 i \hat \sigma_2 \sin \theta \sin \theta',
\end{eqnarray*}
Similar to $E_{1g}$ case for non-zero $\delta_s$, there is one distinct term in the t-matrix with the coefficient, $\alpha$. It indicates a skew scattering effect which spontaneously distinguishes the particles and the holes. Plugging these functions back into the recurrence relation (\ref{t1aK}) and matching the terms with respect to momentum direction and $\hat \tau_i \hat \sigma_j$ matrix forms, we obtain the coefficients of $\hat t_0^R$ as
\begin{eqnarray}
\hat{t}^{R}_{00} &= & \frac{-1}{\pi N_f} \frac{\cot \delta_0 + i \tilde \gamma_{0} \hat \tau_3}{C_{0}}, \quad C_{0} = \cot^2 \delta_0 + (\tilde\gamma_{0})^2,\\
\hat{t}^{R}_{cc} &= & \frac{-1}{\pi N_f} \frac{\cot \delta_c + i \tilde \gamma_{c} \hat \tau_3}{C_{c}}, \quad C_{c} = \cot^2 \delta_c + (\tilde\gamma_{c})^2,\\
\hat{t}^{R}_{ssc} &= & \frac{-3}{\pi N_f} \frac{\cot \delta_s + i \tilde \gamma_{s} \hat \tau_3}{C_{s} }, \quad C_{s} = \cot^2 \delta_s + (\tilde\gamma_{s})^2,\\
\alpha_0 &=&\frac{\tilde \gamma^1 }{C_{s2}},\quad \alpha = \frac{(\tilde{\gamma}^1)^2}{C_{s2}}, \quad C_{s2} = C_{s}+(\tilde{\gamma}^1)^2.
\end{eqnarray}
In addition, the Gfnc averages are same with \ref{selfConsis} except $\tilde{\gamma}^1 = \frac{3}{2} \langle \frac{\tilde{\Delta}_0 \sin^4 \theta}{D^R} \rangle$. The self-energy is found as
\begin{equation}\label{sigma0RE2g}
\hat \sigma^R_0 (\hat k)= \pi N_f \Gamma_u \Big[ \hat{t}^{R}_{00}+ \hat{t}^{R}_{cc} \cos^2 \theta + \hat{t}^{R}_{ssc} \sin^2 \theta \left(1-\alpha/2 \right) - \frac{3\alpha_0}{2 \pi N_f} e^{i 2 \phi \hat \tau_3}\hat \tau_1 i \hat \sigma_2 \sin^2 \theta \Big] 
\end{equation}

For simplicity, we neglected the complex valued self-energy contribution to $\tilde{\Delta}^R_0$ self-consistently, and only keep the scaling part as $\tilde{\Delta}^R_0 = \Delta_0/d$, where $d = \lvert 1 + \frac{9}{4} \Gamma_u \frac{\tilde{\langle \sin^4 \theta / D^R \rangle}}{2 C_{s2}} \rvert^2$. Hence, $\tilde{\Delta}_0$ becomes identical for the retarded and the advanced part. Using $\hat \sigma^R_0 (\hat k)$ in Eq.\ref{sigma0RE2g}, another self-consistency relation of the diagonal elements is also found to be, 
\begin{equation}\label{selfConsE2g}
\varepsilon^R = \varepsilon - \pi N_f \Gamma_u \Big[ (\hat{t}^{R}_{00})_{30}+ (\hat{t}^{R}_{cc})_{30} \cos^2 \theta + (\hat{t}^{R}_{ssc})_{30} \sin^2 \theta \left(1-\alpha/2 \right)\Big]
\end{equation}

The density of states for $E_{2g}$ is $\rho(\cos{\theta},\varepsilon) = - \frac{1}{\pi} \Im{(\hat g_0^R)_{30}}$ are given as the two right figures of Fig.\ref{ModifiedDOS}. The first plot is the clean limit DOS while the second plot is for the impurity concentration $\Gamma_u = 0.4$ T$_c$. The phase shifts are equal to each other $\delta_0 = \delta_s = \delta_c = \pi/3$ and $T = 0.4$ T$_c$. In the modified DOS, the second plot, there are new BQs along various momentum directions certain momentum directions and it again gives non-zero contribution to thermal conductance even at low temperatures.

\subsubsection{E2g, Anomalous t-matrix and the self-energy}\label{Supp332}
$\hat t_{1a}^K$ vanishes in the polar scattering limit ($v_s,\delta_s \to 0$), $v(\hat k, \hat k') = v_0 + 3 v_c \cos \theta \cos \theta'$, with the zeroth order t-matrix $\hat t_{0}^{R,A} = \hat t_{00}^{R,A} + \hat t_{cc}^{R,A} \cos \theta \cos \theta'$.
\begin{eqnarray}
\hat{\sigma}^K_{1a} &\sim& \langle \hat{t}_0^R \hat{g}_{1a}^{K,ns} \hat{t}_0^A \rangle \\
&\sim& \big[\left(\hat{t}_0^R\right)_{even} \langle \left(\cdot\right)_{even} \sin \theta^{''} \cos \theta^{''} \cos \phi'' \rangle \nonumber + \langle \left(\cdot \cdot\right)_{even} \sin \theta^{''} \cos \theta^{''} \cos \phi'' \rangle\left(\hat{t}_0^A\right)_{even} \big] \sin \theta \nonumber\\
&=& 0. \nonumber
\end{eqnarray}

Both averages disappear and therefore the anomalous Keldsyh self-energy is zero, $\hat{g}^{K,V}_{1a} \to 0$. The components of $(\cdot)_{even}$ and $(\cdot\cdot)_{even}$ have either $m=0,2$ angular momentum components meanwhile the multiplicative integrand has $m = 1$ component.

\textbf{Below, we consider only $v_0,v_s$}. The effective scattering potential is, $v(\hat k, \hat k') = v_0 + 3 v_s \sin \theta \sin \theta' \cos{(\phi - \phi')}$. In this limit, ($\delta_c \to 0$), the components of the $k_z$-channel in the t-matrix vanishes, $\hat{t}^{R}_{cc} \to 0$. The form of $\hat{g}_{1a}^{K,ns}$ is given in Eq.\ref{g1aknsExplicit}. The only difference is the order parameter and the momentum dependence of the modified particle-hole energies, $\varepsilon^{R,A}(\theta)$.

- Step 1:
Inserting $\hat g_{1a}^{K,ns}$ into the anomalous t-matrix equation, the anomalous self-energy is obtained.
\begin{eqnarray}
t_{1a}^{K,ns} &=& \pi N_f\langle \hat t^R_{0} \frac{\hat g_{1a}^{K,ns}}{\pi} \hat t^A_{0}\rangle,\\
&=& \pi N_f t^R_{0,00} \langle \frac{\hat g_{1a}^{K,ns}}{\pi} (\hat t^A_{0})_{odd} \rangle + \pi N_f \langle (t_0^R)_{odd} \frac{\hat g_{1a}^{K,ns}}{\pi} \rangle \hat t^A_{0,00},\\
\hat\sigma_{1a}^{K,ns}(\hat k) &=& \Gamma_u \left[ \hat P \frac{k_x}{k_f} + \hat S \frac{k_y}{k_f} + \hat G \frac{k_x+ i k_y \tau_3}{k_f} \hat \tau_1 i \hat \sigma_2 \right] b(\varepsilon,T).
\end{eqnarray}
Note that $\hat P = P_0 + P_3 \hat \tau_3, \hat S=S_0 + S_3 \hat \tau_3,\hat G = G_0 + i G_3 \hat \tau_3$ are diagonal matrix coefficients. The general symmetry considerations of Keldysh Gfncs and the self energies constraints these matrices as $\underline{ P}^\intercal = [P_0 , P_3,S_0 ,S_3,G_0,G_3] \in \mathbb{R}^6$.
The relation between each of the coefficients and the integral averages, $\underline{ \Omega}^\intercal = [\Omega_0 , \Omega_3,\beta_0 ,\beta_3]$ are
\begin{eqnarray}
\begin{bmatrix} 
\hat P\\ \\\hat S \\ \\ \hat G \end{bmatrix} 
&=& \frac{1}{2} \begin{bmatrix}
\left(\hat t_{S0^+}^{+} (1 - \Re{\alpha}) - i \hat t_{S0^+}^{-} \Im{\alpha}\right) \hat{\Omega} -3 \pi N_f \left[ \Re{\alpha_0} \left( \Re{(\hat t_{00}^{R})} \beta_0 - \Im{(\hat t_{00}^{R})} \beta_3 \tau_3 \right)+\Im{\alpha_0} \left( \Im{(\hat t_{00}^{R})} \beta_0 + \Re{(\hat t_{00}^{R})} \beta_3 \tau_3 \right) \right]\\
\\
\left(-i \hat t_{S0^-}^{+} \Re{\alpha}+ \hat t_{S0^+}^{+} \Im{\alpha}\right) \hat{\Omega}\tau_3 -3 \pi N_f \left[ \Re{\alpha_0} \left( \Im{(\hat t_{00}^{R})} \beta_0 + \Re{(\hat t_{00}^{R})} \beta_3 \tau_3 \right)-\Im{\alpha_0} \left( \Re{(\hat t_{00}^{R})} \beta_0 - \Im{(\hat t_{00}^{R})} \beta_3 \tau_3 \right) \right]\tau_3\\
\\
\left[\hat t^{-}_{S0^+} (-0.5 + \Re{\alpha}) + i \hat t_{S0^-}^{-}\Im{\alpha}\right] \hat{\beta} -3 \pi N_f \left[ \Re{\alpha_0} \left(\hat t_{00}^{+} \Omega_0 +\hat t_{00}^{-} \Omega_3 \tau_3 \right)-i \Im{\alpha_0} \left(\hat t_{00}^{-} \Omega_0 +\hat t_{00}^{+} \Omega_3 \tau_3 \right) \right] 
\end{bmatrix} \nonumber \\ .
\end{eqnarray}
Note, $\hat t_{00}^{\pm} = (\hat t_{00}^{R}\pm\hat t_{00}^{R,*-})$ and the matrix $\hat t_{00}^{R,-}$ indicates the same matrix $\hat t_{00}^{R}$ with negative $\tau_3$ component. The integrals defined in $E_{1g}$ section, which are given in Eq.\ref{integralBeta}, $\beta_0(\varepsilon),\beta_3(\varepsilon)$ are modified by the replacement $\cos \theta \to \sin \theta$ as there is no $\cos \theta$ term in $E_{2g}$ order parameter. Moreover, $\hat \beta = \beta_0 + \beta_3 \tau_3$ and $\hat{\Omega} = \Omega_0 + \Omega_3 \tau_3$ integrals can be group into two as follow,
\begin{equation}\label{E2gLowTIntegrals}
\begin{bmatrix}\Omega_0(\varepsilon) \\ \beta_3(\varepsilon) \end{bmatrix} = \langle \frac{2 \Re{D^R}}{D_1} \begin{bmatrix}1 + \frac{\lvert \varepsilon^R \rvert^2-\lvert \Delta(\hat k_1) \rvert^2}{\lvert D^R \rvert^2}\\\frac{2\Im{\varepsilon^R} \Delta_0 }{\lvert D^R \rvert^2 }\sin^2 \theta \end{bmatrix} \sin^2 \theta_{1} \rangle_{\theta_{1}},
\end{equation}
is categorized as the first group with finite values at low temperatures. Also,
\begin{equation}\label{E2gFiniteTIntegrals}
\begin{bmatrix}\Omega_3(\varepsilon) \\ \beta_0(\varepsilon) \end{bmatrix} = \langle \frac{4 \Im{(\sigma^R_0)_{00}}}{D_1}\begin{bmatrix}\Im{\frac{\varepsilon^R}{D
^R}} \frac{2 \Re{D^R}}{D_1} \\ \Im{\frac{\Delta_0}{D^R}} \sin^2 \theta_{1} \end{bmatrix}\sin^2 \theta_{1}\rangle_{\theta_{1}},
\end{equation}
is the second group that dominates at finite temperatures.

Finally, the matrices $\hat t^{\pm}_{S0^{(\pm)}}$ are abbreviations for $ \hat t^{\pm}_{S0^{(\pm)}}\equiv (\pi N_f)^2\left(\hat{t}^R_{ssc} \hat{t}^{A\pm}_{00} (\pm) \hat{t}^R_{00} \hat{t}^{A\pm}_{ssc}\right)/2$,
\begin{eqnarray*}
\hat t^{+}_{S0^+}/(\pi N_f)^2 
&=& \Re{\left[ t_{ssc}^{R} t_{00}^{R*} \right]},\\
&=& \Re{\left[ (t_{ssc}^{R})_{00}(t_{00}^{R*})_{00} + (t_{ssc}^{R})_{30}(t_{00}^{R*})_{30} \right]} + \Re{\left[ (t_{ssc}^{R})_{00}(t_{00}^{R*})_{30} + (t_{ssc}^{R})_{30}(t_{00}^{R*})_{00} \right]} \tau_3,\\
\hat t^{+}_{S0^- }/(\pi N_f)^2 
&=& i \Im{\left[ t_{ssc}^{R} t_{00}^{R*} \right]},\\
&=& i \Im{\left[ (t_{ssc}^{R})_{00}(t_{00}^{R*})_{00} + (t_{ssc}^{R})_{30}(t_{00}^{R*})_{30} \right]} + i \Im{\left[ (t_{ssc}^{R})_{00}(t_{00}^{R*})_{30} + (t_{ssc}^{R})_{30}(t_{00}^{R*})_{00} \right]} \tau_3,\\
\hat t^{-}_{S0^{+}}/(\pi N_f)^2 
&=& \Re{\left[ (t_{ssc}^{R})_{00}(t_{00}^{R*})_{00} - (t_{ssc}^{R})_{30}(t_{00}^{R*})_{30} \right]} - i \Im{\left[ (t_{ssc}^{R})_{00}(t_{00}^{R*})_{30} - (t_{ssc}^{R})_{30}(t_{00}^{R*})_{00} \right]} \tau_3,\\
\hat t^{-}_{S0^{-}}/(\pi N_f)^2 
&=& i \Im{\left[ (t_{ssc}^{R})_{00}(t_{00}^{R*})_{00} - (t_{ssc}^{R})_{30}(t_{00}^{R*})_{30} \right]} - \Re{\left[ (t_{ssc}^{R})_{00}(t_{00}^{R*})_{30} - (t_{ssc}^{R})_{30}(t_{00}^{R*})_{00} \right]} \tau_3.
\end{eqnarray*}

\begin{figure}
\includegraphics[width=0.90
\textwidth]{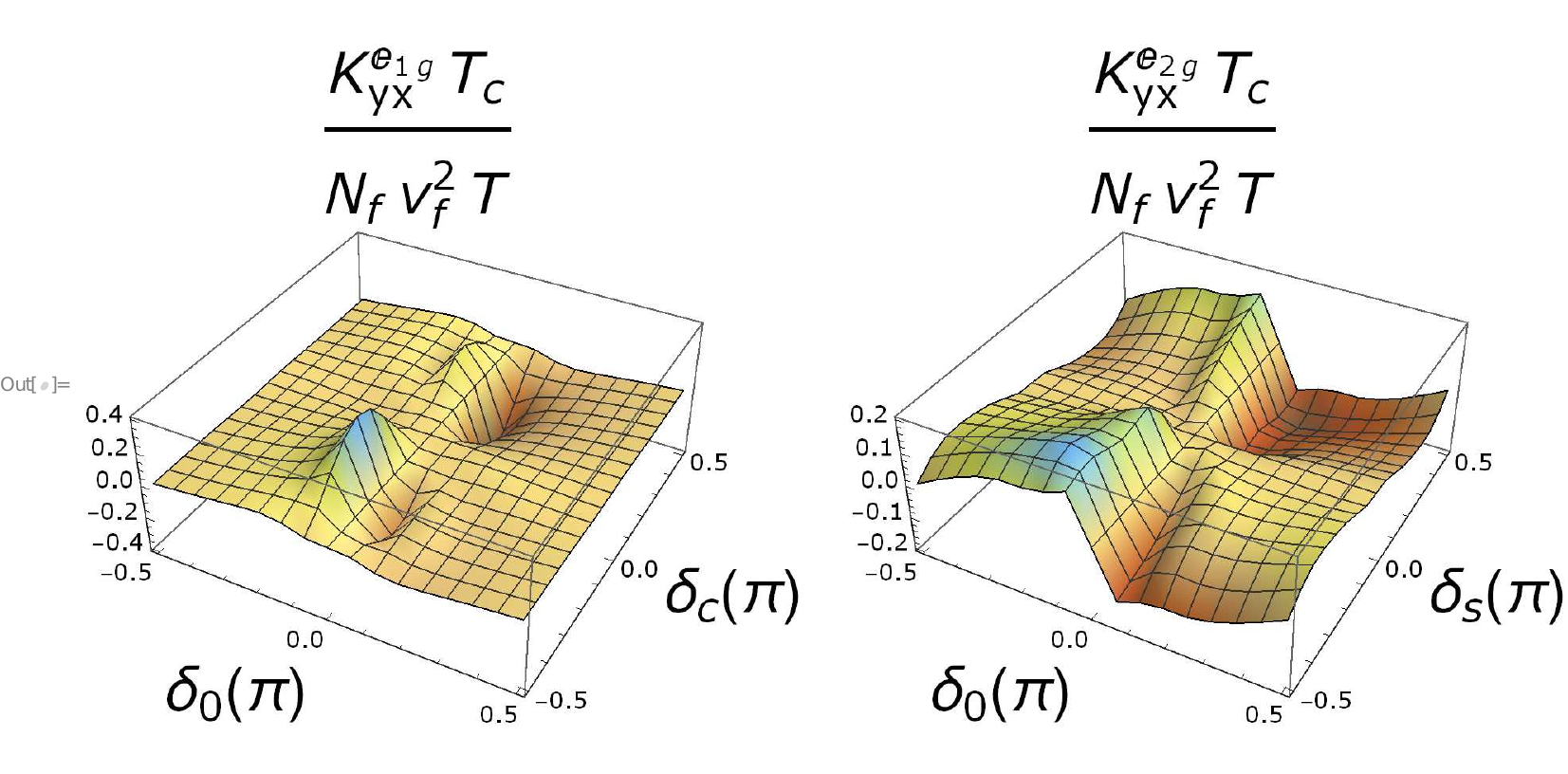}
\caption{Thermal Hall conductance $\kappa_{yx}$ (in units of $\frac{N_f v^2_f T}{T_c}$) in the phase shift space $\delta_0 - \delta_c$ for $E_{1g}$ on the left plot and in the phase shift space $\delta_0 - \delta_s$ for $E_{2g}$ on the right plot for $\Gamma_u = 0.04 T_c$, $T=0.3 T_c$. The inversion of the phase shifts, $\{ \delta_0, \delta_s\} \to \{- \delta_0,- \delta_s\}$ effectively means the particle-hole transformation, and as can be seen, $\kappa_{yx}$ is reversed in both cases, as expected.}
\label{Kyxd0dsE2gE1g}
\end{figure}

- Step 2: Plugging $\hat\sigma_{1a}^{K,ns}(\hat k)$ into $\hat g_{1a}^{K,V}(\hat k)$, the explicit form of $\hat g_{1a}^{K,V}(\hat k)$ is found to be,
\begin{equation}
\hat g_{1a}^{K,V}(\hat k) = \Gamma_u \Big[ \hat X_c \frac{k_x}{k_f} + \hat X_s \frac{ k_y }{k_f}+ \hat{Y} \frac{k_x+ i k_y \tau_3}{k_f} \hat \tau_1 i \hat \sigma_2 \Big] b(\varepsilon,T).
\end{equation}
where the coefficients $\underline{X}^\intercal = [X_{c,0} ,X_{c,3},X_{s,0} ,X_{s,3},Y_{1,0},Y_{1,3} ] \in \mathbb{R}^6$ are related to $\underline{P}$, $\underline{X} = \underline{\underline{W}}^X \underline{P}$,
\begin{eqnarray}\label{g1aKVCoeffE2g}
\begin{bmatrix}X_{c,0} \\\\X_{s,0}\\\\X_{c,3} \\\\X_{s,3} \\\\Y_{0}\\ \\Y_{3} \end{bmatrix} 
&=& \scriptscriptstyle\begin{bmatrix} \frac{2 \Re{D^R}}{D_1}\left[ \left(1+\frac{\lvert \varepsilon^R\rvert^2 - \lvert \tilde \Delta(\hat k) \rvert^2}{\lvert D^R \rvert^2}\right) P_0 - \frac{2\Im{(\varepsilon^R)} \tilde{\Delta}_0 \sin^2 \theta}{\lvert D^R \rvert^2}G_3 \right] +\frac{4\Im{\sigma_{00}^{R}}}{D_1} \Im{\frac{\varepsilon^R}{D^R}}P_3 \\ \\
\frac{2 \Re{D^R}}{D_1}\left[ \left(1+\frac{\lvert \varepsilon^R\rvert^2 - \lvert \tilde \Delta(\hat k) \rvert^2}{\lvert D^R \rvert^2}\right) S_0 - \frac{2\Im{(\varepsilon^R)} \tilde{\Delta}_0 \sin^2 \theta}{\lvert D^R \rvert^2}G_0 \right] +\frac{4\Im{\sigma_{00}^{R}}}{D_1} \Im{\frac{\varepsilon^R}{D^R}}S_3 \\ \\
\frac{2 \Re{D^R}}{D_1}\left[ \left(1+\frac{\lvert \varepsilon^R\rvert^2 + \lvert \tilde \Delta(\hat k) \rvert^2}{\lvert D^R \rvert^2}\right) P_3 + \frac{2\Re{(\varepsilon^R)} \tilde{\Delta}_0\sin^2 \theta}{\lvert D^R \rvert^2}G_0 \right] +\frac{4\Im{\sigma_{00}^{R}}}{D_1} \Im{\frac{\varepsilon^R}{D^R}}\left[ P_0 - \Re{\frac{\tilde \Delta_0 \sin^2 \theta}{D^R}}G_3 \right] \\ \\
\frac{2 \Re{D^R}}{D_1}\left[ \left(1+\frac{\lvert \varepsilon^R\rvert^2 + \lvert \tilde \Delta(\hat k) \rvert^2}{\lvert D^R \rvert^2}\right) S_3 + \frac{2\Re{(\varepsilon^R)} \tilde{\Delta}_0 \sin^2 \theta}{\lvert D^R \rvert^2}G_3 \right] +\frac{4\Im{\sigma_{00}^{R}}}{D_1} \Im{\frac{\varepsilon^R}{D^R}}\left[ S_0 + \Re{\frac{\tilde \Delta_0 \sin^2 \theta}{D^R}}G_0 \right] \\ \\
\frac{2 \Re{D^R}}{D_1}\left( \left[ 1- \lvert \frac{\varepsilon^R}{D^R} \rvert^2 \right] G_0 - \frac{ \tilde{\Delta}_0 \sin^2 \theta}{\lvert D^R \rvert^2}\left[ \Re{\varepsilon^R} P_3+ \Im{\varepsilon^R}S_0\right] \right)
+\frac{4 \Im{\sigma_{00}^R}}{D_1} \left( \Re{(\frac{\varepsilon^R}{D^R})}G_3 - \frac{\tilde \Delta_0 \sin^2 \theta}{2}\left[ \Im{\frac{1}{D^R}} P_0 - \Re{\frac{1}{D^R}} S_3\right] \right)\\ \\
\frac{2 \Re{D^R}}{D_1}\left( \left[ 1- \lvert \frac{\varepsilon^R}{D^R} \rvert^2 \right] G_3 + \frac{ \tilde{\Delta}_0 \sin^2 \theta}{\lvert D^R \rvert^2}\left[ \Re{\varepsilon^R} S_3- \Im{\varepsilon^R}P_0\right] \right)
+\frac{4 \Im{\sigma_{00}^R}}{D_1} \left(-\Re{(\frac{\varepsilon^R}{D^R})}G_0 + \frac{\tilde \Delta_0 \sin^2 \theta}{2}\left[ \Im{\frac{1}{D^R}} S_0 + \Re{\frac{1}{D^R}} P_3\right] \right)
\end{bmatrix} \nonumber \\
\end{eqnarray}

The coefficients of the vertex correction Keldysh Gfnc is obtained. However, the coefficients are renormalized if the problem is treated with full self-consistency, $\underline{X} \to \underline{\tilde{X}}$. For full self-consistency, we first replace $\underline{P} \to \underline{\tilde{P}}$ in the above relation and secondly replace $\hat g_{1a}^{K,ns}(\hat k) \to \hat g_{1a}^{K,V}(\hat k)$ in Eq.\ref{t1aK}. The anomalous t-matrix relation becomes, 
\begin{equation}
\hat t_{1a}^{K} - \hat t_{1a}^{K,ns} = \Big[ (\hat t^R_{0})_{00} \langle \frac{\hat g_{1a}^{K,V}}{\pi} (\hat t^A_{0})_{odd} \rangle + \langle (\hat t_0^R)_{odd} \frac{\hat g_{1a}^{K,V}}{\pi} \rangle (\hat t^A_{0})_{00} \Big].
\end{equation}
Repeating the same procedure retains a self-consistency relation for the renormalized coefficients of $\hat \sigma_{1a}^K$, $\underline{\tilde{P}}$ by connecting the renormalized $\underline{X}$ and $\underline{\tilde{P}}$ with the relation, $\underline{\tilde{X}} = \underline{\underline{\mathcal{W}}}^{XP} \underline{\tilde{P}}$. Together, we obtain the following relation,
\begin{eqnarray}
\underline{\tilde{P}} - \underline{P} &=& \Gamma_u \langle \sin^2 \theta \underline{\underline{\mathcal{W}}}^{PX} \underline{\underline{\mathcal{W}}}^{XP} \underline{\tilde{P}} \rangle,\\
\underline{\tilde{P}} &=& \left( 1- \Gamma_u \underline{\underline{\mathcal{W}}}^{PX} \langle \sin^2 \theta \underline{\underline{\mathcal{W}}}^{XP}\rangle \right)^{-1} \underline{P}
\end{eqnarray}

At low temperatures, $T\to 0$, the excitations are only possible in the vicinity of the nodes (poles) where $\varepsilon \to 0$. Then, $\varepsilon^R \to i (\gamma_0 + \gamma_s \sin^2 \theta)$, and $\gamma_0$ and $ \gamma_s$ are the bandwidths with positive definite values. The two of the integral averages vanish in this limit, $\Omega_3=\beta_0 \to 0$. Non-self consistent contribution to the longitudinal conductivity is $\frac{\kappa^{ns}_{xx}T_c}{\frac{\pi^2}{3}N_f v_f^2 T}= \Omega_0$. The non-zero integrals, $\Omega_0$ and $\beta_3$ are evaluated as $\Omega_0(0) = \langle \frac{ \gamma^2(\theta) }{D^3}\sin^2 \theta \rangle \sim \frac{\gamma_0}{\Delta^2_0}$, $\beta_3(0) = \langle \frac{ \gamma(\theta) \Delta_0}{D^3}\sin^4 \theta \rangle \sim \gamma_0/\Delta^2_0 \ln{\frac{\tilde{\Delta}_0}{\gamma_0}}$ and $\tilde{\gamma}^1 \approx constant$. The vertex corrections to THCs, $\kappa^V_{xx},\kappa^V_{yx}$ are found to be
\begin{equation}
\label{kyxE2gT0Supp}
\frac{\begin{bmatrix}\kappa^V_{xx}\\ \kappa^V_{yx}\end{bmatrix} T_c}{\frac{\pi^2}{3}N_f v_f^2 T} \approx \Gamma_u T_c \frac{3/4}{ C_0 C_s} \begin{bmatrix}(1-\Re{\alpha}) \left( \cot{\delta_0}\cot{\delta_s}+\tilde{\gamma}_0\tilde{\gamma}_s \right) \Omega_0^2 -\Re{\alpha} \left(\cot{\delta_0}\cot{\delta_s}-\tilde{\gamma}_0\tilde{\gamma}_s \right) \beta_3^2 - 2\frac{\tilde{\gamma}_0 \tilde{\gamma}_s C_s}{C_{s2}}\Omega_0 \beta_3 \\ \left(\cot{\delta_0} \tilde{\gamma_s}+\cot{\delta_s} \tilde{\gamma_0}\right)(1+\Re{\alpha}) \beta_3^2 + \Re{\alpha} \left(\cot{\delta_0} \tilde{\gamma_s}- \cot{\delta_s} \tilde{\gamma_0}\right) \Omega_0^2 + \Re{\alpha_0} \cot{\delta_0} \Omega_0 \beta_3 C_s \end{bmatrix} .
\end{equation}
\end{widetext}
In addition, $\tilde{\gamma}_0 = \langle \frac{\gamma}{D} \rangle \sim \frac{\gamma_0}{\tilde{\Delta_0}} \ln{\frac{\tilde{\Delta_0}}{\gamma_0}}$, $\tilde{\gamma}_s = \langle \frac{\gamma \sin^2 \theta}{D} \rangle \sim \frac{\gamma_0}{\tilde{\Delta_0}}$ and $\tilde{\gamma}^1 = \langle \frac{\tilde{\Delta}_0 \sin^4 \theta}{D} \rangle \sim \textit{constant}$, where $\Re{\alpha_0}\sim\Re{\alpha} \sim \frac{1}{C_{s2}} = (\cot^2 \delta_s + \frac{\gamma^2_0}{\tilde{\Delta^2_0}})^{-1}$. 

Eq.\ref{kyxE2gT0Supp} is a complicated expression, but the low temperature integrals are estimated in orders of magnitudes above. For typical, non-vanishing and non-diverging values of $ \cot \delta_0$ and $ \cot \delta_s$, the terms with $\beta_3^2 \cot \delta_0$ or $\Omega_0^2 \cot \delta_0$ has the dominant contribution for $\kappa_{xx}$ while it is the mixed term, $\cot{\delta_0} \Omega_0 \beta_3$ for $\kappa_{yx}$. Evaluating the vertex corrections with the dominant terms along with $\Gamma_u/C_0 = \gamma_0/\tilde{\gamma}_0$, we obtain the following expression,
\begin{equation}\label{kyxE2gT0approxSupp}
\sim \frac{T_c}{\tilde{\Delta}_0} \frac{\gamma_0^2 \ln{\frac{\tilde{\Delta}_0}{ \gamma_0}}}{\tilde{\Delta}^2_0}\Re{\alpha}  \begin{bmatrix} - \frac{\cot \delta_s}{C_s} \\ 1\end{bmatrix} \cot \delta_0
\end{equation}

At finite temperatures, the thermal Hall conductance is dominated by $\beta_0^2$ or $\Omega_3^2$ (with the vertex correction it is complicated mixing of all averages) due to the non-zero $\Im{(\sigma_0^R)_{00}}$ term. Physically, if the bare Gfncs are to be used, the imaginary parts of $(\Sigma^{R,A})_{00}$ change the lifetime of electrons and holes. As an overall, the lifetime for electrons and holes differ, $\tau_{e,h}^{-1} \propto - \Im{\left[(\Sigma^{R})_{30}\pm(\Sigma^{R})_{00}\right]}$. This effect is not visible directly in quasiclassical approach as  $\hat g^{R,A}$ and  $(\sigma^{R,A})_{00}$ commute in Eq.\ref{transportEq} and $\hat g^{R,A}_0$ does not include $\tau_0 \sigma_0$ component. Interestingly, it still modifies the non-equilibrium occupation because $(\sigma^{R,A})_{00}$ are explicitly present in Eq.\ref{NR}.

\section{$\kappa_{yx}$ dependence on the impurity concentration, $\Gamma_u$}\label{Supp4}
In literature, up to our knowledge, there is no discussion on the $\kappa_{yx}$ vs. $\Gamma_u$ dependence. At finite temperatures, $\kappa_{yx}$ in the clean limit diverges as the typical scattering lifetime $1/\tau \to 0$. In this limit, $\kappa_{xx}^{ns} \sim \int_{\varepsilon}(..) \Omega_0(\varepsilon)$ in Eq.\ref{JxnsE1gSupp}, and expression in the paragraph above Eq.\ref{kyxE2gT0Supp} and $ \Omega_0(\varepsilon)$ also diverges with the same trend $\Omega_0 \sim 1/\Gamma_u$. We therefore examine the ratio $\kappa_{yx}/\kappa_{xx}$.
\begin{eqnarray}
\label{GammaUZeroLimE1g}
\kappa_{yx}^{E_{1g}} &\sim& \Gamma_u \int_{\varepsilon}(..) \beta^2_0(\varepsilon), \quad \beta_0(\varepsilon) \sim \frac{1}{\Gamma_u},
\\
\label{GammaUZeroLimE2g}
\kappa_{yx}^{E_{2g} } &\sim& \Gamma_u \int_{\varepsilon}(..) \Omega^2_0(\varepsilon), \quad \Omega_0(\varepsilon) \sim \frac{1}{\Gamma_u},
\\
\label{kappaYXOverkappaXX}
\kappa_{yx}/\kappa_{xx} &\sim& constant.
\end{eqnarray}

\begin{figure}
\includegraphics[width=0.40
\textwidth]{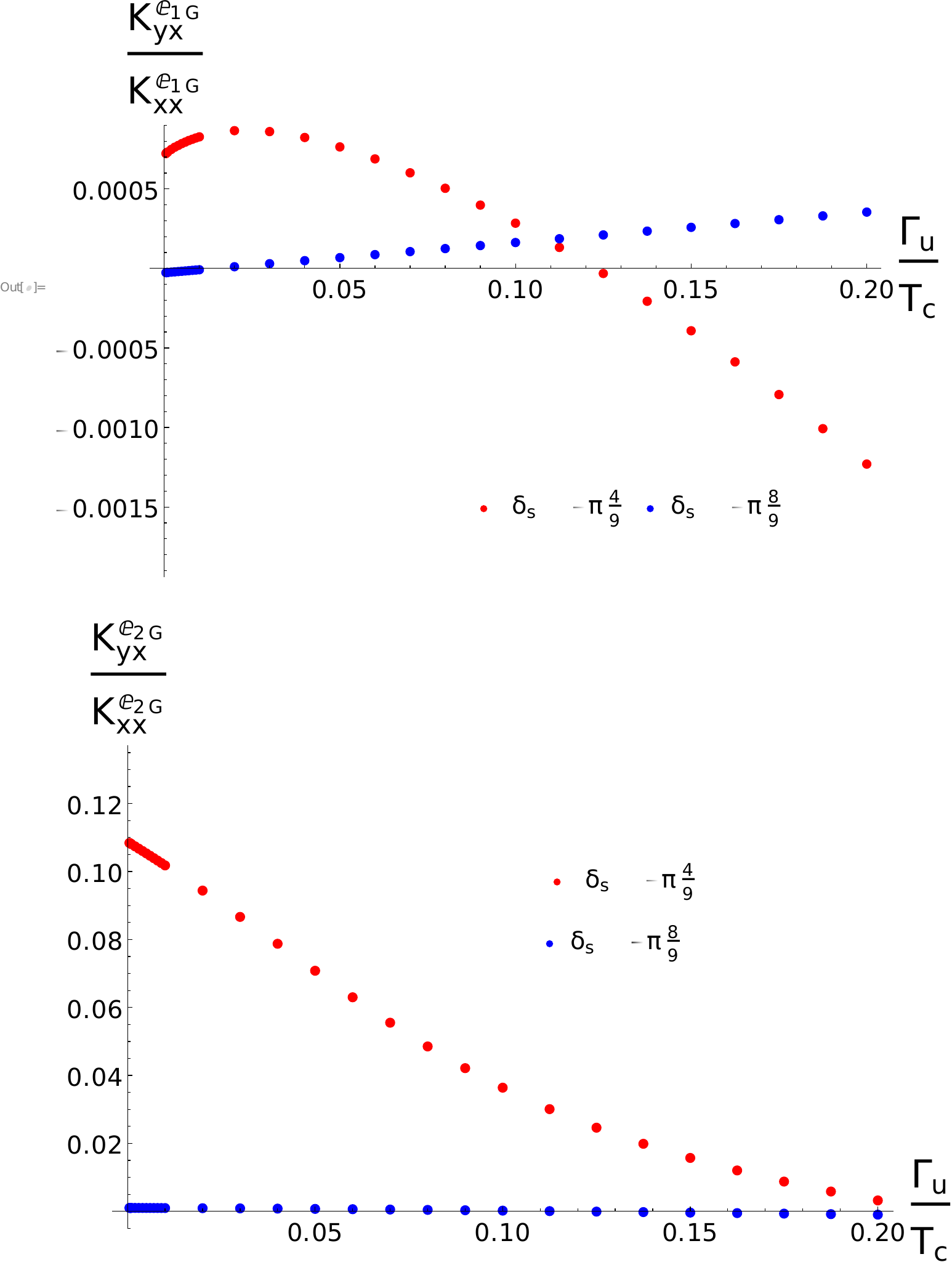}
\caption{$\kappa_{yx}(T)/\kappa_{xx}(T)$ vs. $\Gamma_u/T_c$, Thermal Hall conductance normalized to the longitudinal conductance as a function of the impurity concentration in units of the superconducting critical temperature. $T^{E_{1g}}=0.35 T_c$ and $T^{E_{2g}}=0.65 T_c$. The presence of $\kappa_{xx}$ avoids a possible divergence occurring in $\Gamma_u \to 0$ limit.}
\label{KyxGammaU}
\end{figure}
The expression $(..)$ is the rest of the uninteresting terms that determine the finite numerical scale in Eq.\ref{Jiy}. Note that the integrals have the same form for $E_{1g}$ and $E_{2g}$, one should consult with the subsequent subsection for the explicit functions. 
At small concentrations, the impurity contribution dominates over the topological part due to the longer scattering lifetime ($\tau$) for the Bogoliubov quasiparticles. However, in the extremely clean limit, ballistic regime will be reached and the approach of the present paper does not apply. Our discussion in this section assumes that the ballistic regime has not been reached.

Quantitatively, in Fig.\ref{KyxGammaU}, we calculate the thermal Hall conductivity normalized to $\kappa_{xx}$ as a function of the impurity concentration, $\Gamma_u$ (in units of $T_c$). The immediate observation validates our claim that suggests the decrease in $\kappa_{yx}/\kappa_{xx}$ as a function of $\Gamma_u$ up to the physical values of $\Gamma_u$ where the superconducting phase is not suppressed by the impurity scattering \citep{graf1996thermal,joynt1997bound}. $\kappa_{yx}$ can change sign as the different contributions overcome at different impurity concentrations, though we neglect the suppression of superconductivity for large $\Gamma_u$ values in our approach. It should also be noted that $\kappa_{yx}$ is also dependent on the phase shifts and it can be suppressed at all $\Gamma_u$ values as seen on the right plot for $E_{2g}$ case in Fig.\ref{KyxGammaU}. In summary, the impurity contribution, $\kappa_{yx}^{IM}$ typically dominates over $\kappa_{yx}^{topo.}$ since the topological contribution is independent of $\Gamma_u$.
For completeness, let us show the low concentration impurity limit for the integrals. We present the results in terms of the BQ lifetime, $\tau^{-1} = \Gamma_u \frac{\cot{\delta_0}^2 + \lvert \tilde{\gamma}_0 \rvert^2}{\lvert \cot{\delta_0}^2 + ( \tilde{\gamma}_0 )^2 \rvert^2} \Re{ \tilde{\gamma}_0}$. For $\Gamma_u \to 0$, $\tau \to \infty$. We omitted anisotropic part of the impurity scattering for clearer forms as they do not change the relevant scales. Note that $x = \frac{\varepsilon}{\Delta_0}$,
\begin{eqnarray}
\label{omega0}
\Omega_0 (x) &\to& \begin{cases} 
\text{constant} , & \lvert x \rvert < 1,
\\
\tau \langle \frac{\sqrt{x^2 - f^2(\hat k)} }{x} \frac{(k_x^2 + k_y^2)}{k_f^2} \rangle , & \lvert x \rvert > 1. 
\end{cases}
\\
\label{beta0}
\beta_0(x) &\to& \begin{cases} 
\tau h(\varepsilon) \langle \frac{\sqrt{x^2 - f^2(\hat k)}f^2(\hat k) }{x^2 + (x^2 - f^2(\hat k)) h^2(\varepsilon)} \rangle_{x^2 > f^2} , & \lvert x \rvert <1, 
\\
\text{constant} , & \lvert x \rvert > 1.
\end{cases} \nonumber \\
\\
\label{beta3}
\beta_3(x) &\to& \text{constant}
\end{eqnarray}
The numerical values for the averages changes for each order parameter and it is denoted by $f^2(\hat k)$. $f^2(\hat k)^{E_{1g}}= k_z^2 (k_x^2 + k_y^2)/k_f^4$ and $f^2(\hat k)^{E_{2g}}= (k_x^2 + k_y^2)^2/k_f^4$. Also, $h(\varepsilon)$ is independent of $\hat k$ and has an explicit dependence on the phase shifts, $h(\varepsilon) = \frac{2 \cot \delta_0}{\cot^2 \delta_0 + \lvert \tilde{\gamma}_0 \rvert^2} \Im{ \langle \frac{- i x}{\sqrt{ f^2(\hat k)-x^2 }} \rangle_{x^2<f^2} }$.
\end{document}